\documentclass[pre,twocolumn,floatfix,showpacs]{revtex4-1}
\usepackage{graphicx}
\usepackage{txfonts}

\begin{document}
\title{Corrugation instabilities of the Riemann problem in relativistic hydrodynamics}

\date{}

\author{Patryk Mach}
\affiliation{M. Smoluchowski Institute of Physics, Jagiellonian University, Reymonta 4, 30-059 Krak\'ow, Poland}

\begin{abstract}
Corrugation instabilities occurring for solutions of the Riemann problem in relativistic hydrodynamics in which the fluid moves with a non-zero velocity tangent to the initial discontinuity are studied numerically. We perform simulations both for ultrarelativistic and perfect gas equations of state. We focus on a set of problems with moderately relativistic velocities but exhausting all possible wave patterns of solutions. Perturbations are applied to the shape of the initial discontinuity. Instabilities that develop are only restricted to a region around a contact discontinuity. Both shock and rarefaction waves appear to be stable.
\end{abstract}


\maketitle

\section{Introduction}
The term Riemann problem usually refers to an initial value formulation for a hyperbolic set of partial differential equations where initial data consist of two constant states separated by the discontinuity in the form of a plane surface.

Although more than 150 years have passed since the publication of Riemann's original paper \cite{riemann}, the significance of the Riemann problem in relativistic hydrodynamics has been truly recognized only recently (see \cite{marti_mueller_rev} for an introduction).

In this paper we are concerned with the general case of the Riemann problem in relativistic hydrodynamics, where the fluid is allowed to move with the velocity tangent to the initial discontinuity. In Newtonian hydrodynamics such solutions are not qualitatively different from those where the gas flows only in the direction normal to this surface. In the relativistic case, however, all velocities couple through Lorentz factors, and the resulting wave pattern of the solution depends on the velocities tangent to the initial discontinuity. Relativistic solutions of the Riemann problem with non-vanishing tangential velocities were first obtained numerically by Pons, Mart\'{\i} and M\"{u}ller in 2000 for the perfect gas equation of state \cite{pons_marti_mueller}. Some solutions for other equations of state were also computed in \cite{giacomazzo_rezzolla, meliani_et_al}, and an analytic solution for the ultrarelativistic equation of state was presented by Pi\c{e}tka and the author in \cite{mach_pietka}.

Solutions of this type are usually used in order to construct and test three dimensional hydrodynamical codes. However in 2002 Aloy and Rezzolla discovered a hydrodynamical boosting mechanism accelerating gas in relativistic astrophysical jets to large Lorentz factors \cite{aloy_rezzolla}. The analysis of this mechanism was based on solutions of the relativistic hydrodynamical Riemann problem with non-zero velocities tangent to the initial discontinuity. It is thus clear that the role of this class of Riemann problems is not restricted to numerical methods.

This paper is devoted to three dimensional numerical studies of corrugation instabilities occurring in solutions of the Riemann problem with non-zero tangent velocities both for ultrarelativistic and perfect gas equations of state. Such instabilities, mainly of the Kelvin--Helmholtz type, were suggested e.g.~in \cite{aloy_rezzolla}. Since solutions of the hydrodynamical Riemann problem usually consist of different elementary waves: rarefaction and shock waves, a contact discontinuity and some constant states, the question of their stability is a little bit subtle. It makes sense to ask whether all parts of the solution become unstable or not. An instability of the region around a contact discontinuity does not have to imply that the surface of a shock wave present in the solution has to be unstable. Also, when performing studies of the stability of solutions, one has to decide on a certain notion of stability, and the final answer may depend on this choice. We will address these issues in Sec.~V of this paper.

We restrict ourselves to problems with moderately relativistic fluid velocities ($v \sim 1/2$), and concentrate on solutions characterized by different wave patterns, i.e., with two shock waves, two rarefaction waves or a shock and a rarefaction wave respectively. The question of stability of solutions with ultrarelativistic speeds is still open.

There exist many numerical codes solving equations of relativistic hydrodynamics for the perfect gas equation of state, or for general equations of state where the pressure can be expressed in terms of the rest-mass density and the specific internal energy. The case with ultrarelativistic equation of state is different in this respect; it basically requires a new formulation of the numerical scheme. Such a numerical code has been written for the purpose of this work and will also be described here.

The order of this paper is as follows. In the next section we collect all basic equations necessary to explain our approach. Sec.~III discusses the structure of exact solutions of the Riemann problem in relativistic hydrodynamics. In Sec.~IV we describe new elements of numerical methods used in this paper. Sec.~V contains numerical results on the stability of the Riemann problem in relativistic hydrodynamics. A short summary can be found in Sec.~VI.

\section{Basic equations} 

Conservation of the energy and the momentum in relativistic hydrodynamics can be expressed as
\begin{equation}
\label{energy_momentum_cons}
\partial_\mu T^{\mu \nu} = 0,
\end{equation}
where $T^{\mu \nu}$ denotes the perfect fluid energy-momentum tensor
\begin{equation}
T^{\mu \nu} = (p + \rho) u^\mu u^\nu + p \eta^{\mu \nu}.
\end{equation}
Here $p$ is the pressure, $\rho$ the energy density, $u^\mu$ denote components of the four-velocity of the fluid, and $\eta^{\mu\nu}$ are the components of the Minkowski metric. In the above formulae Greek indices range from 0 to 3 and the convention with
\[ \eta^{\mu \nu} = \mathrm{diag} (-1,+1,+1,+1) \]
is assumed. If the equation of state is in the barotropic form $p = p(\rho)$, the above equations constitute a complete set of equations of hydrodynamics. The ultrarelativistic equation of state of the form $p = c_s^2 \rho$, with $c_s^2$ being a constant (the local speed of sound) is a good example here.

More generally one introduces the rest-mass density $n$, so that the continuity equation
\begin{equation}
\label{rest_mass_cons}
\partial_\mu \left( n u^\mu \right) = 0
\end{equation}
is satisfied, and the equation of state is expressed in terms of the rest-mass density $n$ and the specific internal energy $\epsilon$, defined as $\epsilon = (\rho - n)/n$. This is the case of the perfect gas equation of state: $p = (\gamma - 1)n \epsilon$, where $\gamma$ is a constant. In such a case the complete set of equations of hydrodynamics consists of four Eqs.~(\ref{energy_momentum_cons}) and Eq.~(\ref{rest_mass_cons}).

For the purpose of evolutionary problems, it is convenient to rewrite Eqs.~(\ref{energy_momentum_cons}) and Eq.~(\ref{rest_mass_cons}) in the form, where the derivatives with respect to time and space are separated explicitly. We introduce the Lorentz factor $W = u^0$ and components of the three-velocity $v^i = u^i/W$. Eqs.~(\ref{energy_momentum_cons}) can be now written as
\begin{equation}
\label{u_and_f}
\partial_t \mathbf U + \partial_i \mathbf F^i = 0,
\end{equation}
where
\begin{eqnarray}
\label{u_ultra}
\mathbf U & = & \left( (\rho + p) W^2 - p, (\rho + p) W^2 v^1, \right. \nonumber \\
& & \left. (\rho + p) W^2 v^2, (\rho + p) W^2 v^3 \right)^T,
\end{eqnarray}
\begin{eqnarray}
\label{f_ultra}
\mathbf F^i & = & \left( (\rho + p) W^2 v^i, (\rho + p) W^2 v^i v^1 + \delta^{i 1} p, \right. \nonumber \\
& & \left. (\rho + p) W^2 v^i v^2 + \delta^{i 2}p, (\rho + p) W^2 v^i v^3 + \delta^{i 3} p  \right)^T,
\end{eqnarray}
and $\delta^{ij}$ denotes the Kronecker's delta.

The set of Eqs.~(\ref{energy_momentum_cons}) and Eq.~(\ref{rest_mass_cons}) can be also expressed in the form of Eq.~(\ref{u_and_f}). To this end, it is customary to introduce the specific enthalpy $h =  (\rho + p)/n = 1 + \epsilon + p/n$. Vectors $\mathbf U$ and $\mathbf F^i$ must be now 5-dimensional, namely
\begin{equation}
\label{u_perfect}
\mathbf U = \left( n h W^2 - p, n h W^2 v^1, n h W^2 v^2, n h W^2 v^3, n W \right)^T
\end{equation}
and
\begin{eqnarray}
\label{f_perfect}
\mathbf F^i & = & \left( n h W^2 v^i, n h W^2 v^i v^1 + \delta^{i 1} p, \right. \nonumber \\
& & \left. n h W^2 v^i v^2 + \delta^{i 2}p, n h W^2 v^i v^3 + \delta^{i 3} p, n W v^i  \right)^T. 
\end{eqnarray}

In this paper we will refer to Eqs.~(\ref{u_and_f}) with $\mathbf U$ and $\mathbf F^i$ given by (\ref{u_ultra}) and (\ref{f_ultra}) and the ultra-relativistic equation of state as to system I. Eqs.~(\ref{u_and_f}) with $\mathbf U$ and $\mathbf F^i$ defined as in (\ref{u_perfect}) and (\ref{f_perfect}) and the perfect gas equation of state will be called system II.

\section{Solutions of the Riemann problem}

Solutions of the Riemann problem for system I were obtained by Smoller and Temple for the case without velocities tangent to the initial discontinuity in \cite{smoller_temple}. Solutions valid for arbitrary velocities (also tangent to the discontinuity) were presented by Pi\c{e}tka and the author in \cite{mach_pietka}.

Analogous solutions for system II were first calculated by Mart\'{\i} and M\"{u}ller for the case without tangential velocities \cite{marti_mueller} and promoted later to the case with non-zero tangential velocities by Pons, Mart\'{\i} and M\"{u}ller in \cite{pons_marti_mueller}.

In order to describe these solutions briefly, we will introduce Cartesian coordinates $x$, $y$, $z$, and assume that the initial discontinuity, dividing two initial states $L$ and $R$, is a plane given by $x = 0$. Here, traditionally, $L$ refers to the ``left'' state, i.e., for $x<0$; $R$  refers to  the``right'' one, that is for $x>0$. In this case solutions of the Riemann problem are self-similar: they depend on $x$ and time $t$ through a single variable $\xi = x/t$, and they are independent of $y$ and $z$.

In all cases solutions can be constructed from self-similar elementary waves: a smooth part---the so-called rarefaction wave $\mathcal R$ and two possible discontinuities, i.e., a shock wave $\mathcal S$ and the contact discontinuity $\mathcal C$. All three kinds of elementary waves can be separated by some constant intermediate states $L_\ast$ and $R_\ast$. Moreover, it can be shown that the initial state $LR$ decays into one of the following possible wave patterns:
\begin{eqnarray*}
LR & \to & L \mathcal S_\leftarrow L_\ast \mathcal C R_\ast \mathcal S_\rightarrow R,\\
LR & \to & L \mathcal R_\leftarrow L_\ast \mathcal C R_\ast \mathcal R_\rightarrow R,\\
LR & \to & L \mathcal S_\leftarrow L_\ast \mathcal C R_\ast \mathcal R_\rightarrow R,\\
LR & \to & L \mathcal R_\leftarrow L_\ast \mathcal C R_\ast \mathcal S_\rightarrow R,
\end{eqnarray*}
where the arrows refer to the direction in which the waves move with respect to each other (it is customary to refer to these waves as to left and right-moving). By the contact discontinuity we understand a surface across which the pressure and the normal component of the velocity are continuous, while other quantities, like tangent components of the velocity or, in case of system II, the rest-mass density and the specific internal energy, can exhibit a jump. Accordingly, the pressure $p$ and $x$ components of the velocity $v^x$ have to be equal in both intermediate states $L_\ast$ and $R_\ast$. Since it is possible to compute the relation between $p$ and $v^x$ behind the left and right moving waves (either a shock or a rarefaction wave) in terms of the state in front of the wave ($L$ or $R$ depending on the direction of propagation of the wave), we can compute the values of $p$ and $v^x$ in both intermediate states from the condition that these values have to be the same for both right and left moving waves. This calculation identifies the particular type of both waves. Next, it is possible to compute the speeds of propagation of discontinuities and of the front and the tail of the rarefaction wave (if present). At this stage computing all other details of the solution is straightforward. A precise description of this procedure can be found in \cite{pons_marti_mueller, mach_pietka, marti_mueller}.

\section{Description of the numerical codes}

The code used to investigate the instabilities occurring in the solutions of the Riemann problem is based on a Godunov type, high resolution shock capturing scheme (for a textbook exposition see \cite{leveque, laney}). The construction of the code is similar to that described in \cite{aloy_et_al}, and a preliminary version has been presented in \cite{mach}.

The temporal evolution of conserved quantities is implemented according to a variant of method of lines. The space is discretised into cells (zones) labeled by indices $i$, $j$, $k$ in Cartesian directions $x$, $y$ and $z$ respectively. In the following $x_i$, $y_j$ and $z_k$ will denote coordinates of the center of the zone labeled by $i$, $j$ and $k$. Dimensions of the zone will be given by $\Delta x_i$, $\Delta y_i$ and $\Delta z_k$. Positions of the interfaces between zones will be denoted in the usual fashion, where symbol $x_{i+1/2}$ refers to the $x$ coordinate of the interface between zones labeled by $i$ and $i+1$, and positions of the interfaces in directions $y$ and $z$ are denoted in the analogous way. The time derivative of cell-averaged values of $\mathbf U$ is computed according to the following formula
\begin{eqnarray*}
\frac{d \mathbf{U}_{i,j,k}}{dt} & = & - \frac{1}{\Delta x_i} \left( \hat{\mathbf F}^x_{i+1/2, j, k} - \hat{\mathbf F}^x_{i - 1/2, j, k} \right) \\
& & - \frac{1}{\Delta y_j} \left( \hat{\mathbf F}^y_{i, j + 1/2, k} - \hat{\mathbf F}^y_{i, j - 1/2, k} \right) \\
& & - \frac{1}{\Delta z_k} \left( \hat{\mathbf F}^z_{i, j, k + 1/2} - \hat{\mathbf F}^z_{i, j, k - 1/2} \right),
\end{eqnarray*}
where $\hat{\mathbf F}^x_{i+1/2, j, k}$, $\hat{\mathbf F}^x_{i - 1/2, j, k}$, $\hat{\mathbf F}^y_{i, j + 1/2, k}$, $\hat{\mathbf F}^y_{i, j - 1/2, k}$, $\hat{\mathbf F}^z_{i, j, k + 1/2}$ and $\hat{\mathbf F}^z_{i, j, k - 1/2}$ denote numerical fluxes defined at the interfaces between adjacent cells. Values of $\mathbf U_{i,j,k}$ are advanced in time using the standard fourth or second order Runge--Kutta method, i.e., either as
\[ \mathbf U_{i,j,k}^{n+1} = \mathbf U_{i,j,k}^n + \frac{1}{6} \Delta t \left( \mathbf k_1 + 2 \mathbf k_2 + 2 \mathbf k_3 + \mathbf k_4 \right), \]
where
\begin{eqnarray*}
\mathbf k_1 & = & \frac{d \mathbf U_{i,j,k}}{dt} \left( \mathbf U_{i,j,k}^n \right), \\
\mathbf k_2 & = & \frac{d \mathbf U_{i,j,k}}{dt} \left( \mathbf U_{i,j,k}^n + \frac{1}{2} \Delta t \mathbf k_1 \right), \\
\mathbf k_3 & = & \frac{d \mathbf U_{i,j,k}}{dt} \left( \mathbf U_{i,j,k}^n + \frac{1}{2} \Delta t \mathbf k_2 \right), \\
\mathbf k_4 & = & \frac{d \mathbf U_{i,j,k}}{dt} \left( \mathbf U_{i,j,k}^n + \Delta t \mathbf k_3 \right),
\end{eqnarray*}
or according to
\[ \mathbf U_{i,j,k}^{n+1} = \mathbf U_{i,j,k}^n + \Delta t \mathbf k_2. \]
Here the upper index $n$ numbers subsequent time steps of size $\Delta t$. 

Clearly, the key problem is to compute the values of numerical fluxes $\hat{\mathbf F}$. In \cite{mach} we have presented preliminary tests obtained by using relativistic Harten, van Leer, Lax, Einfeldt (HLLE) formulae for system I (for the description of the HLLE solver see \cite{harten_lax_leer, einfeldt, schneider_et_al}). Here we go a little bit further and employ a modified version of a flux formula introduced originally by Donat and Marquina \cite{marquina} and used in \cite{aloy_et_al}.

Suppose we want to compute a numerical flux at the interface between two Riemann states $\mathbf U_L$ and $\mathbf U_R$. Let $\lambda_p$, $\mathbf l_p$ and $\mathbf r_p$ be the eigenvalues, left eigenvectors and right eigenvectors of the Jacobian $\partial \mathbf F^i / \partial \mathbf U$ respectively. Moreover, subscripts $L$ and $R$ will refer to the values obtained for the left and right states $\mathbf U_L$ and $\mathbf U_R$. Numerical fluxes $\hat \mathbf F^i$ are computed as
\begin{eqnarray*}
\hat{\mathbf F}^i & = & \frac{1}{2} \left\{ \mathbf F^i(\mathbf U_L) + \mathbf F^i (\mathbf U_R) \right. \\
& & \left. - \sum_p \mathrm{max}_{L,R}| \lambda_p | \left( (\mathbf l_{p,R} \cdot \mathbf U_R) \mathbf r_{p,R} - (\mathbf l_{p,L} \cdot \mathbf U_L) \mathbf r_{p,L} \right) \right\}.
\end{eqnarray*}

The values of $\lambda_p$, $\mathbf l_p$ and $\mathbf r_p$ should be computed analytically. They are given in \cite{banylus_et_al} for system II, and have to be computed separately for system I. It can be noted, however, that the precise knowledge of all these terms is not necessary. The only required quantities are the eigenvalues $\lambda_p$ and all vectors $(\mathbf l_{p} \cdot \mathbf U) \mathbf r_{p}$.

In general, for a barotropic fluid with $p = p(\rho)$ and $c_s^2 = dp / d\rho$ the eigenvalues of $\partial \mathbf F^i / \partial \mathbf U$ are
\[ \lambda_0 = v^i, \]
\[ \lambda_\pm = \frac{v^i (1 - c^2_s) \pm c_s \sqrt{\left( 1 - v_k v^k \right) \left( 1 - v_k v^k c^2_s - (v^i)^2 (1 - c^2_s) \right)}}{1 - v_k v^k c^2_s}. \]
The eigenvalue $\lambda_0$ is twofold degenerate. The expression for $\lambda_\pm$ can be also written as
\[ \lambda_\pm = \frac{v^i \pm A}{1 \pm v^i A} \]
with $A^{-2} = 1 + W^2 \left( 1 - (v^i)^2 \right)(1 - c^2_s)/c^2_s$. If $v^i$ is the only component of the velocity then $A = c_s$.

The remaining formulae will be given for $\partial \mathbf F^x/\partial \mathbf U$. Expressions for $\partial \mathbf F^y/\partial \mathbf U$ and $\partial \mathbf F^z/\partial \mathbf U$ can be easily obtained from the symmetry. (There is also no reason to treat Jacobians $\partial \mathbf F^i/\partial \mathbf U$ separately in the code. It is enough to implement formulae for $\hat{\mathbf F}^x$; other fluxes can be computed by swapping the order of velocities and conserved momenta.) Let us introduce the square of the tangential velocity $v_t^2 = v_y^2 + v_z^2$, and define a bunch of auxiliary quantities:
\begin{eqnarray*}
\Theta_\pm & = & c_s^2 \rho (1 - v_x^2) \left(v_t^2 - v_x^2 - 1 + 2 \lambda_{\mp} v_x \right), \\
\Sigma_\pm & = & (1 - \lambda_\pm v_x) \left( v_t^2 (1 - v_t^2) + 2 v_t^2 - v_x^2 (1 - v_x^2) \right) \\
& & - 2 v_t^2 (1 - v_x^2), \\
\Omega_\pm & = & \lambda_\pm (1 + c_s^2 v_k v^k) - (1 + c_s^2) v_x, \\
\Delta_\pm & = & \frac{\Omega_\pm \left( \Theta_\pm + p (c_s^2 \Sigma_\pm + (1 - v_k v^k)(\lambda_\mp - v_x)v_x ) \right)}{(\lambda_\mp - \lambda_\pm) \Xi}.
\end{eqnarray*}
In these terms
\begin{eqnarray*}
\sum_{j = 1}^2(\mathbf l_{0,j} \cdot \mathbf U) \mathbf r_{0,j} & = & \frac{p W^2}{1 - v_x^2} \left( 2 v_x v_t^2, v_y (1 - v_x^2 + v_t^2), \right. \\
& & \left. v_z (1 - v_x^2 + v_t^2), 2 v_z^2 \right)^T
\end{eqnarray*}
and
\begin{eqnarray*}
& (\mathbf l_\pm \cdot \mathbf U) \mathbf r_\pm & = \\
& &  \Delta_\pm \left( \frac{(\lambda_\pm - v_x) v_x + c_s^2 \left( 1 - v_t^2 - \lambda_\pm v_x (2 - v_x^2 + v_t^2) \right)}{\lambda_\pm \left( 1 + c_s^2 (v_x^2 - v_t^2) \right) - (1 + c_s^2) v_x}, \right. \\
& & \left. v_y , v_z, \frac{v_x \left( c_s^2 (v_x^2 - v_t^2) - 1 \right) + \lambda_\pm (1 - c_s^2 v_k v^k)}{\lambda_\pm \left( 1 + c_s^2 (v_x^2 - v_t^2) \right) - (1 + c_s^2) v_x} \right)^T.
\end{eqnarray*}

Systems I and II differ also in the numerical procedure used to recover primitive hydrodynamical quantities like $n$, $p$, $v^i$ from the conserved ones. For the perfect gas equation of state such a recovery is performed by means of a Newton--Raphson scheme. For the ultrarelativistic equation of state primitive values can obtained from $\mathbf{U}$ analytically.

The last, key ingredient of the code is the reconstruction procedure used to obtain the states $\mathbf U_L$ and $\mathbf U_R$ based on the values of $\mathbf U$ from the neighboring zones. For the results presented here we have used the Convex Essentially Non Oscillatory (CENO) reconstruction procedure presented in \cite{liu_osher} and applied to the relativistic hydrodynamics by \cite{londrillo_zanna, zanna_bucciantini}.

The general rules of the method can be found in \cite{liu_osher}. Here we will only give final formulae. The CENO reconstruction is applied in each dimension to each component $u$ of the conserved vector $\mathbf U$ separately. Consider the $x$ direction and a zone $x_{i - 1/2} \leq x \leq x_{i + 1/2}$. Let us define
\[ \tilde S_i = \frac{(\Delta x_i^-)^2 u_{i+1} - ((\Delta x_i^+)^2 - (\Delta x_i^-)^2) u_i - (\Delta x_i^+)^2 u_{i-1}}{\Delta x_i^+ \Delta x_i^- (\Delta x_i^+ + \Delta x_i^-)}, \]
\[ \hat S_i = 2 \frac{\Delta x_i^- u_{i+1} - (\Delta x_i^+ - \Delta x_i^-) u_i - \Delta x_i^+ u_{i-1}}{\Delta x_i^+ \Delta x_i^- (\Delta x_i^+ + \Delta x_i^-)}, \]
and
\[ S_i = \mathrm{mm} \left( \frac{u_i+1 - u_i}{\Delta x_i^+}, \tilde S_i, \frac{u_i - u_{i-1}}{\Delta x_i^-} \right), \]
where $\Delta x_i^+ = x_{i+1} - x_i$, $\Delta x_i^- = x_i - x_{i-1}$, and $\mathrm{mm}$ denotes the minmod function, i.e.,
\[ \mathrm{mm} ( x_1, \dots, x_n) = \left\{ \begin{array}{ll} \mathrm{min} \{ x_1, \dots, x_n \}, & \text{if $x_i \geq 0$ for all $i$,}\\ \mathrm{max} \{ x_1, \dots, x_n \}, & \text{if $x_i \leq 0$ for all $i$,}\\ $0$, & \text{otherwise.}  \end{array} \right. \]
Next, we introduce a linear function
\[ L_i (x) = u_i + S_i (x - x_i) \]
and three quadratic polynomials
\[ Q_i^{(k)} = u_{i+k} + \tilde S_{i+k} (x - x_i) + \frac{1}{2} \tilde S_{i+k} (x - x_i)^2, \]
for $k = 0, \pm 1$. Following \cite{liu_osher} we will now choose quadratic polynomials $Q_i^{(k)}$ that are closest to $L(x)$ at $x_{i-1/2}$ and $x_{i+1/2}$ respectively. More precisely, in order to obtain the left state $u^L_{i+1/2}$, we compute differences $d^{(k)} = Q_i^{(k)}(x_{i+1/2}) - L_i(x_{i+1/2})$. In the case where all $d^{(k)}$ have the same sign we choose the polynomial $\tilde Q_i = Q_i^{(k)}$ for which $|d^{(k)}|$ is the smallest. Otherwise we select the linear function $\tilde Q_i = L_i$. The left state $u^L_{i+1/2}$ is now obtained as $u^L_{i+1/2} = \tilde Q_i (x_{i+1/2})$. An analogous selection procedure is applied at the interface $x_{i-1/2}$ in order to compute $u^R_{i-1/2} = \tilde Q_i (x_{i-1/2})$.

The code was tested against different exact solutions of the Riemann problem. The results were slightly better than those presented in \cite{mach}. We have also performed convergence tests for three dimensional runs with satisfactory results, i.e., the solution computed on a coarser grid resembled that computed with a better resolution.

\section{Corrugation instability problem}

\begin{table}[t!]
\caption{Initial data for sample problems that were evolved numerically. Problems (a)--(c) correspond to equations of system I. Problems (d)--(f) apply to system II. The resulting wave pattern is indicated as ``type''. Symbols $\mathcal{SS}$, $\mathcal{RR}$ and $\mathcal{RS}$ refer to configurations with two shock waves, two rarefaction waves and a combination of a shock and a rarefaction wave respectively.}
\begin{ruledtabular}
\begin{tabular}{cccccccccc}
\multicolumn{10}{c}{System I}\\
\hline
Problem & \multicolumn{2}{c}{$\rho_L$} & $v^x_L$ & $v^y_L$ & \multicolumn{2}{c}{$\rho_R$} & $v^x_R$ & $v^y_R$ & Type \\
\hline
(a) & \multicolumn{2}{c}{0.5} &  0.2 &  0.2 & \multicolumn{2}{c}{0.5} & -0.2 & -0.2 & $\mathcal{SS}$ \\
(b) & \multicolumn{2}{c}{0.5} & -0.2 &  0.2 & \multicolumn{2}{c}{0.5} &  0.2 & -0.2 & $\mathcal{RR}$ \\
(c) & \multicolumn{2}{c}{0.5} &  0   &  0.2 & \multicolumn{2}{c}{1.0} &  0   & -0.2 & $\mathcal{RS}$ \\
\hline
\multicolumn{10}{c}{System II}\\
\hline
Problem & $n_L$ & $\epsilon_L$ & $v^x_L$ & $v^y_L$ & $n_R$ & $\epsilon_R$ & $v^x_R$ & $v^y_R$ & Type \\
\hline
(d) & 1.0 & 0.5 &   0.2 &  0.2 & 1.0 & 0.5 & -0.2 & -0.2 & $\mathcal{SS}$ \\
(e) & 1.0 & 0.5 &  -0.2 &  0.2 & 1.0 & 0.5 &  0.2 & -0.2 & $\mathcal{RR}$ \\
(f) & 1.0 & 0.5 &   0   &  0.2 & 1.0 & 1.0 &  0   & -0.2 & $\mathcal{RS}$
\end{tabular}
\end{ruledtabular}
\end{table}

\begin{figure}[t!]
\includegraphics[width=6.6cm]{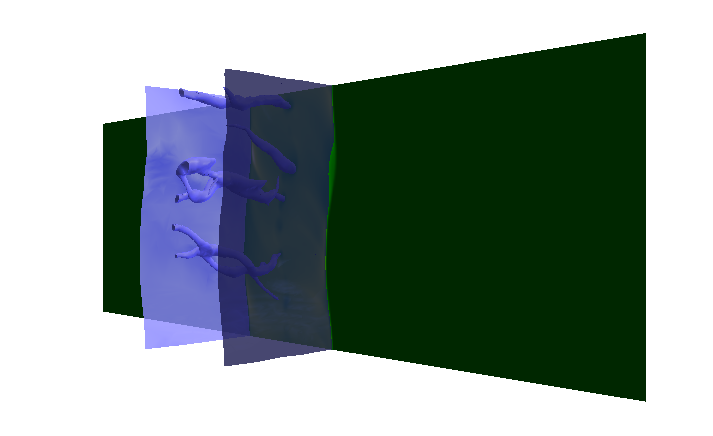}
\includegraphics[width=6.6cm]{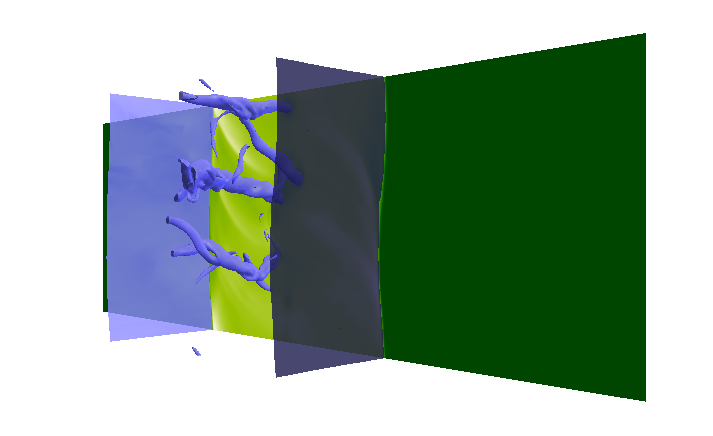}
\includegraphics[width=6.6cm]{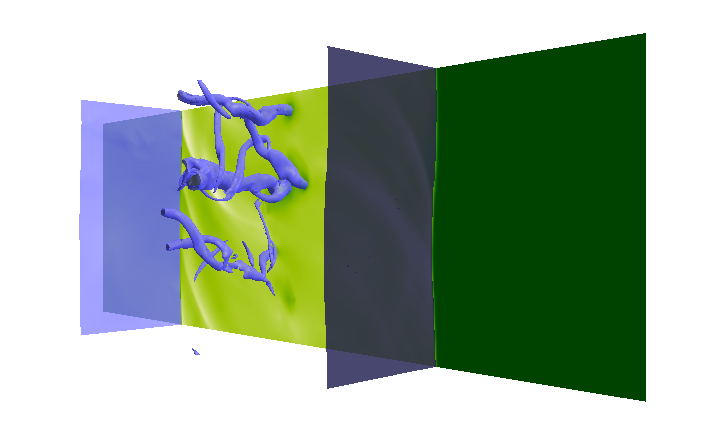}
\includegraphics[width=6.6cm]{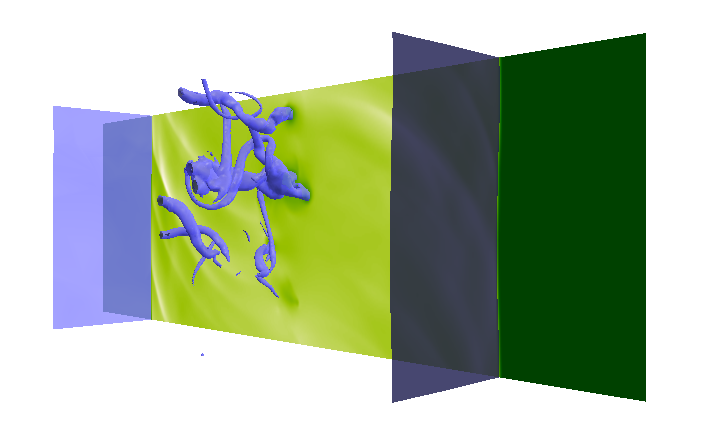}
\includegraphics[width=6.6cm]{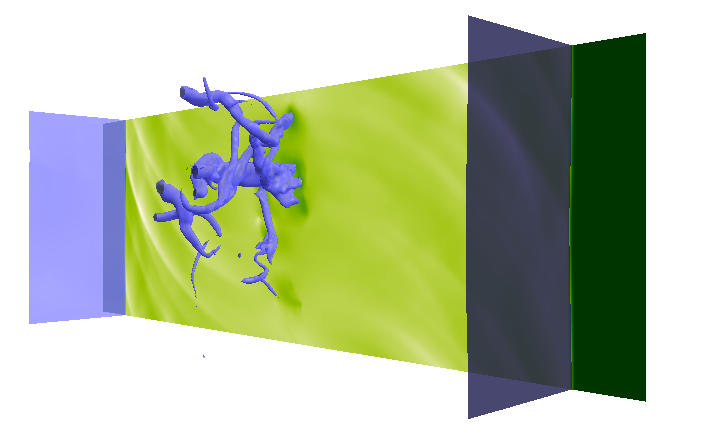}
\caption{Three dimensional distribution of conserved energy density $e$ for problem (a). Subsequent snapshots correspond to evolution times $t = 0.5, 1.0, 1.5, 2, 2.5$.}
\label{prob_a_3d}
\end{figure}

\begin{figure}[t!]
\includegraphics[width=6.6cm]{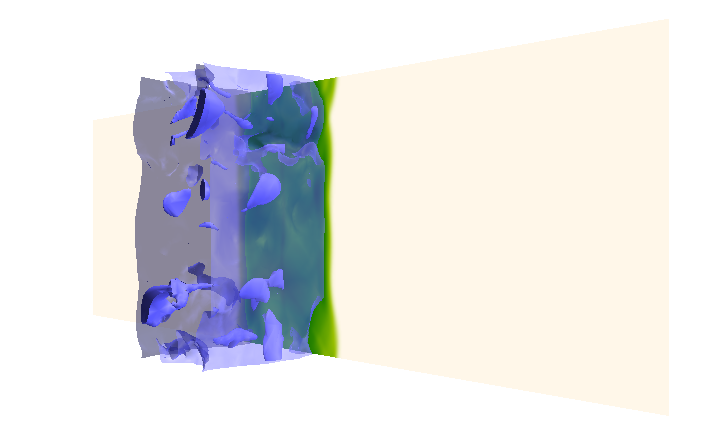}
\includegraphics[width=6.6cm]{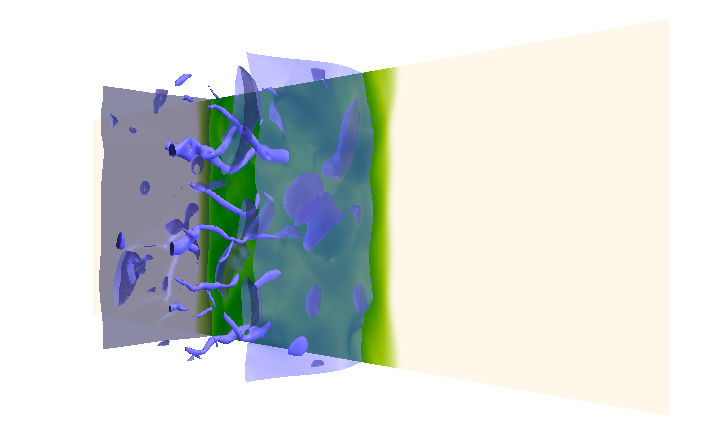}
\includegraphics[width=6.6cm]{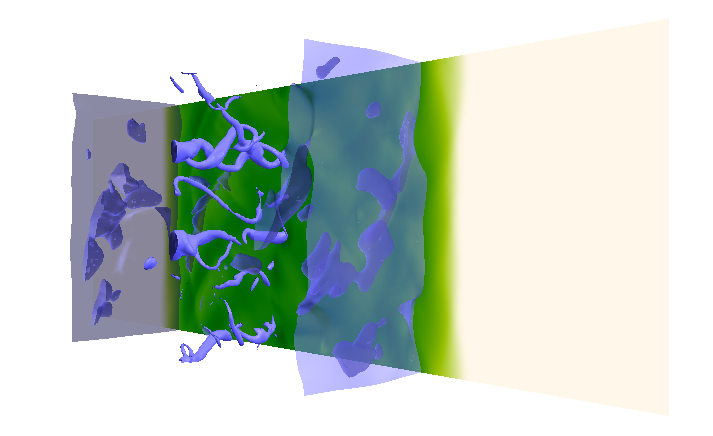}
\includegraphics[width=6.6cm]{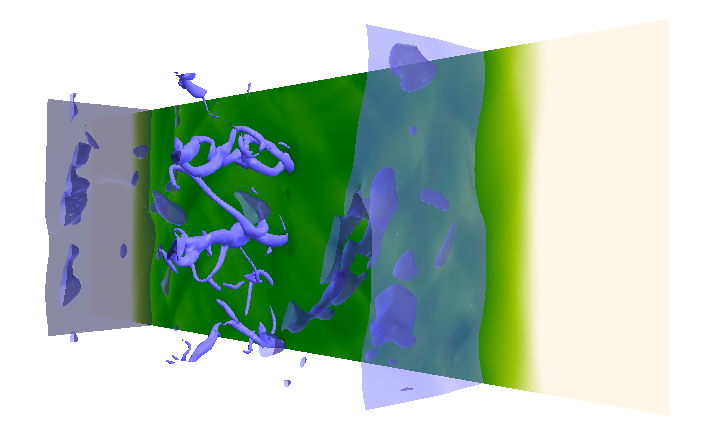}
\includegraphics[width=6.6cm]{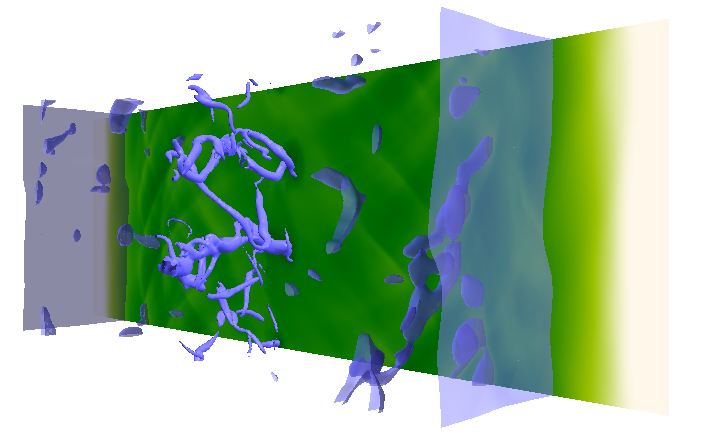}
\caption{Three dimensional distribution of conserved energy density $e$ for problem (b). Subsequent snapshots correspond to evolution times $t = 0.4, 0.8, 1.2, 1.6, 2.0$.}
\label{prob_b_3d}
\end{figure}

\begin{figure}[t!]
\includegraphics[width=6.6cm]{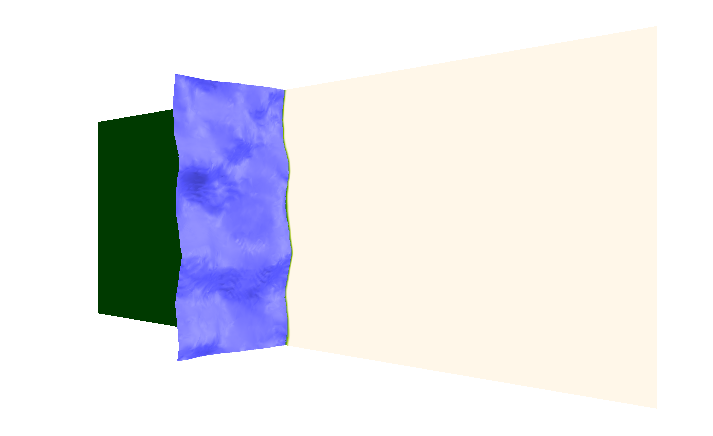}
\includegraphics[width=6.6cm]{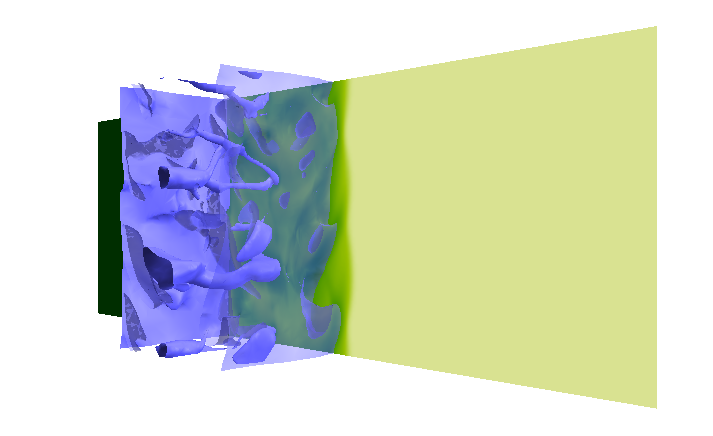}
\includegraphics[width=6.6cm]{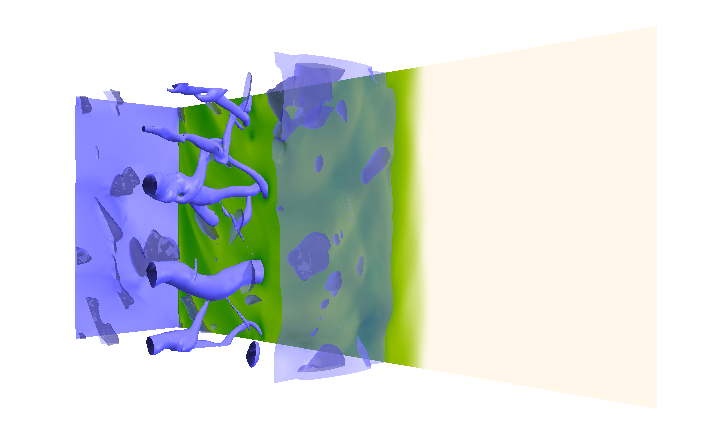}
\includegraphics[width=6.6cm]{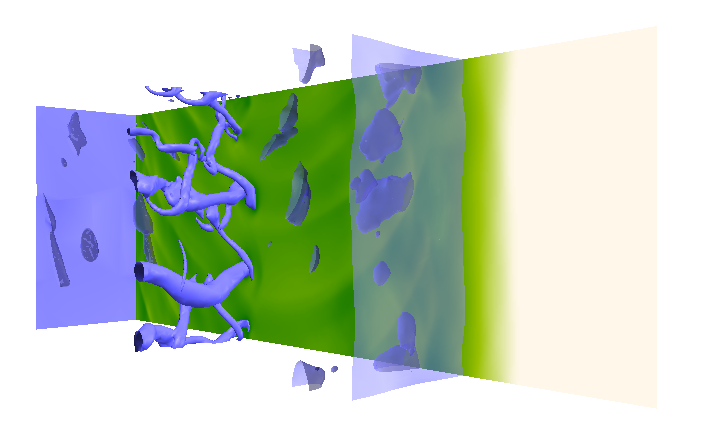}
\includegraphics[width=6.6cm]{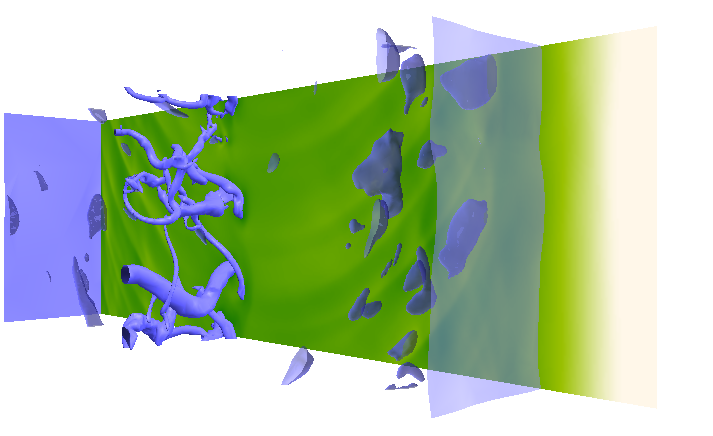}
\caption{Three dimensional distribution of conserved energy density $e$ for problem (c). Subsequent snapshots correspond to evolution times $t = 0.0, 0.6, 1.2, 1.8, 2.4$.}
\label{prob_c_3d}
\end{figure}

\begin{figure}[t!]
\includegraphics[width=6.6cm]{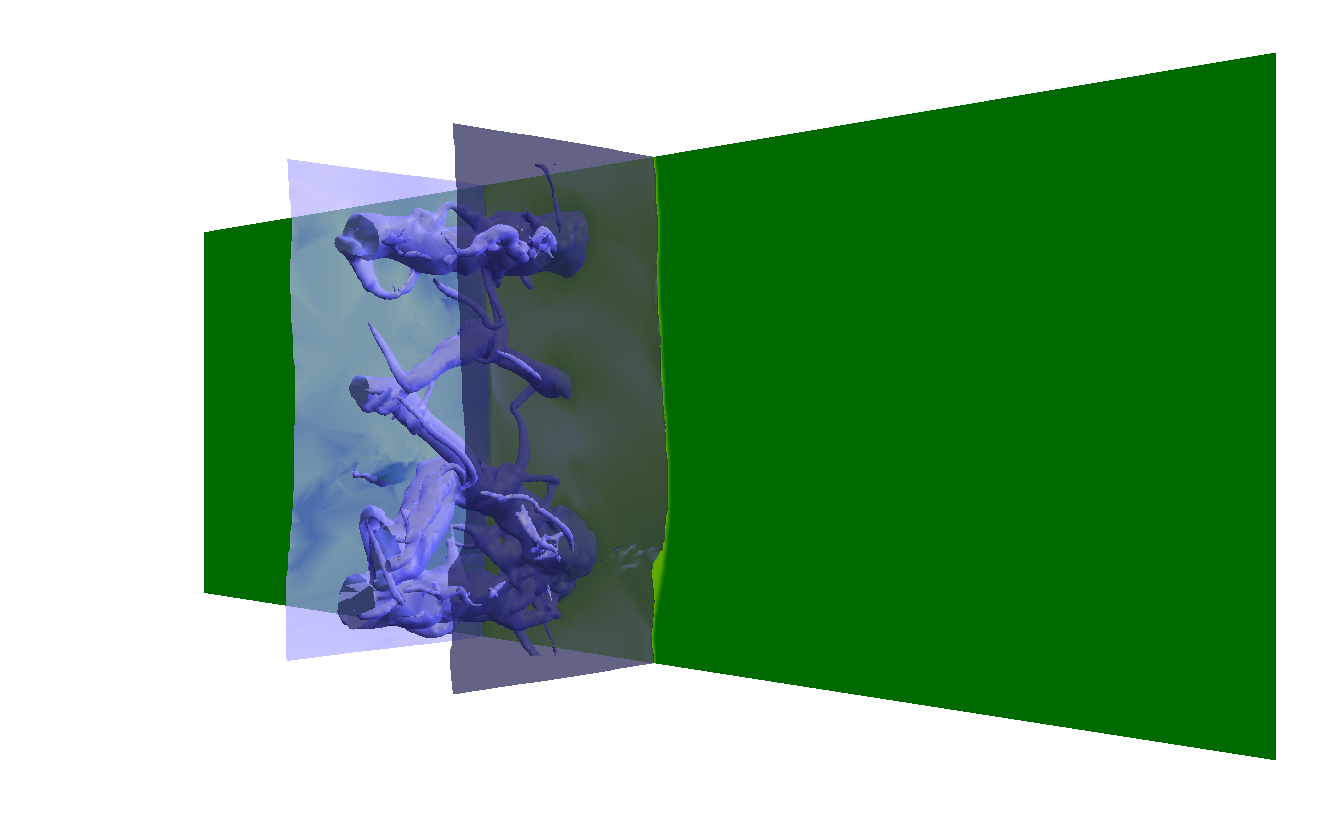}
\includegraphics[width=6.6cm]{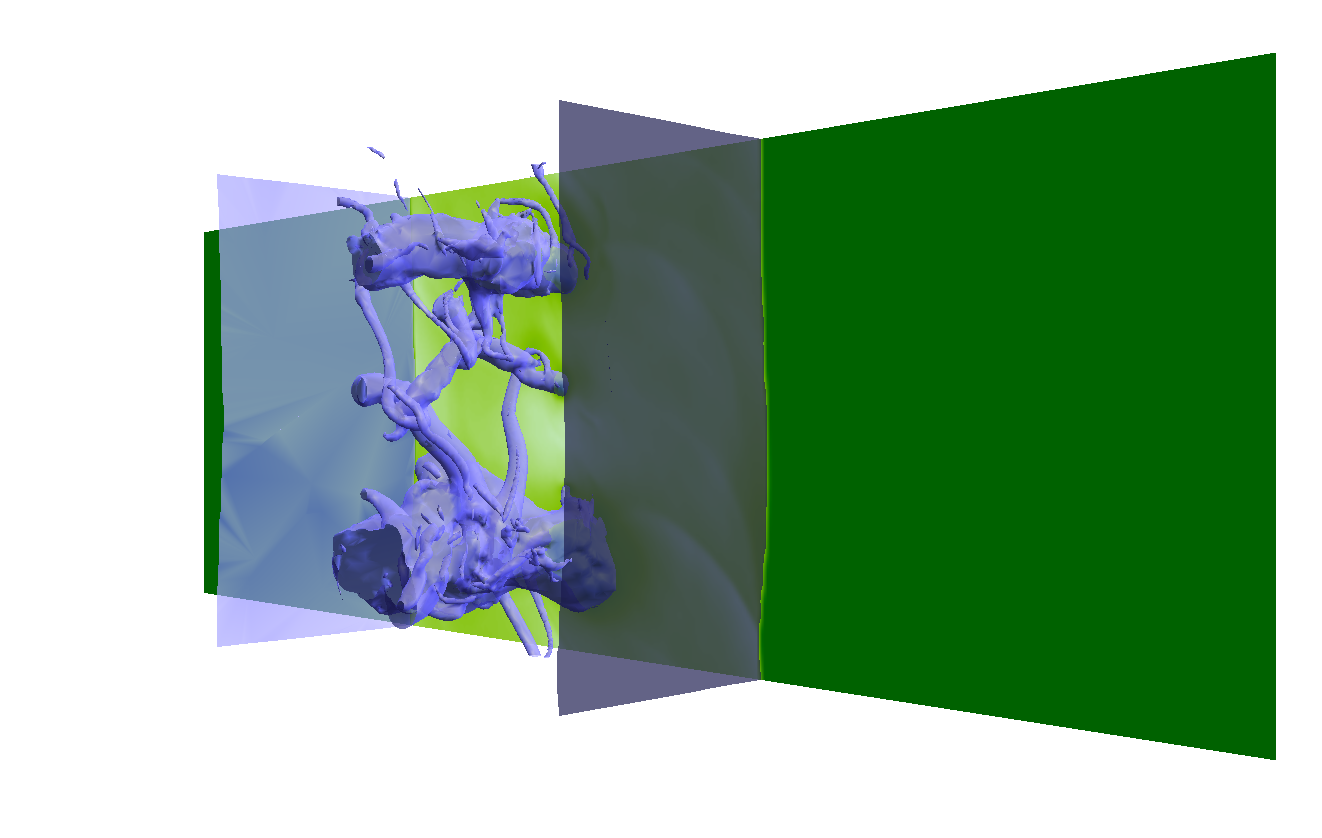}
\includegraphics[width=6.6cm]{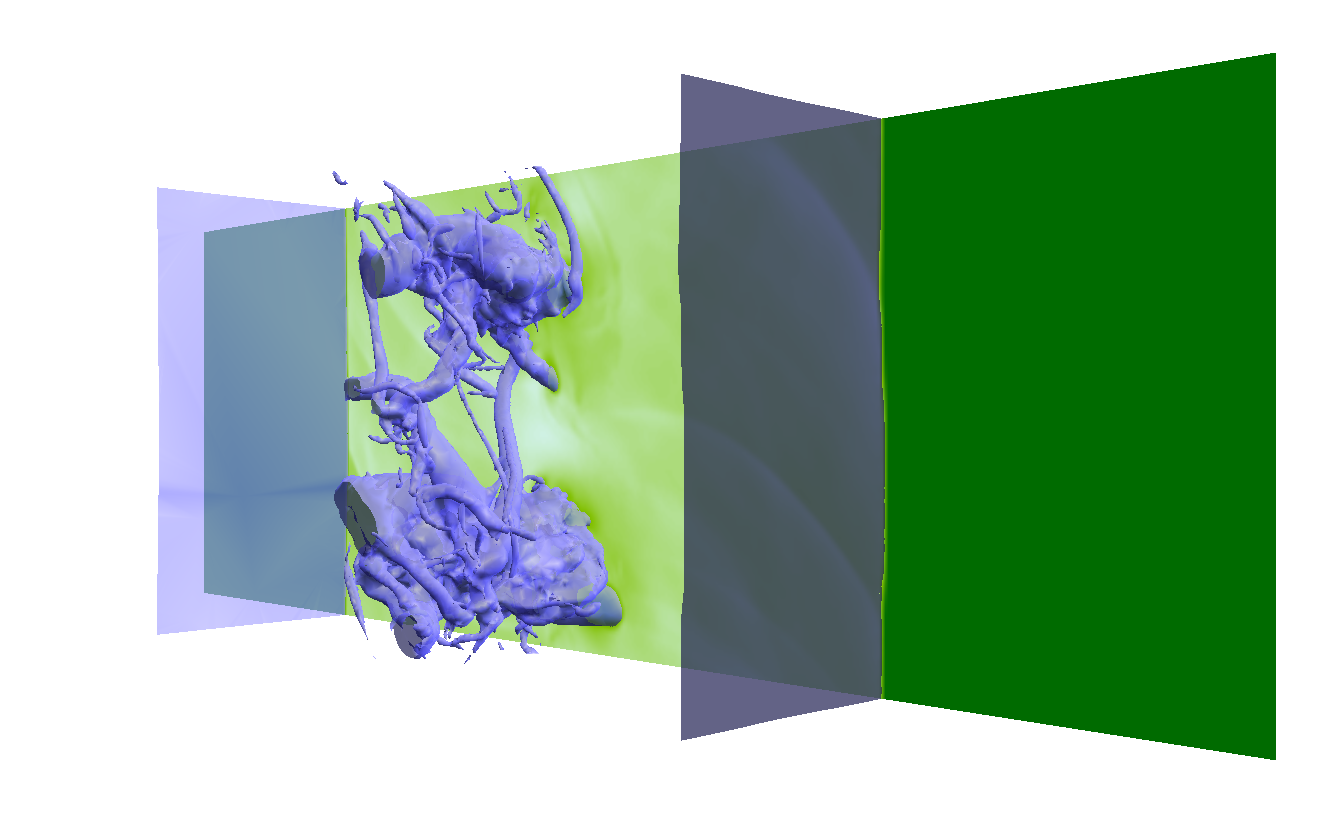}
\includegraphics[width=6.6cm]{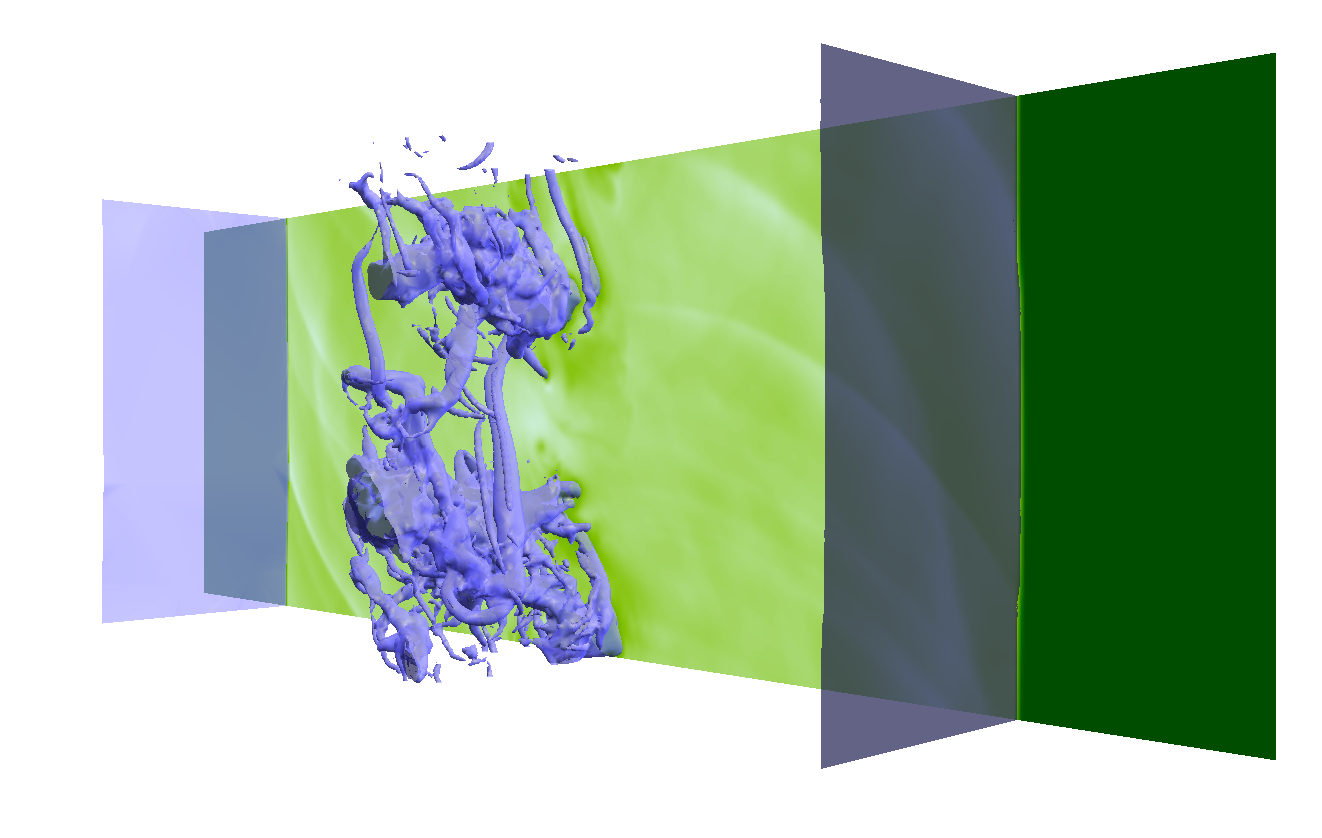}
\includegraphics[width=6.6cm]{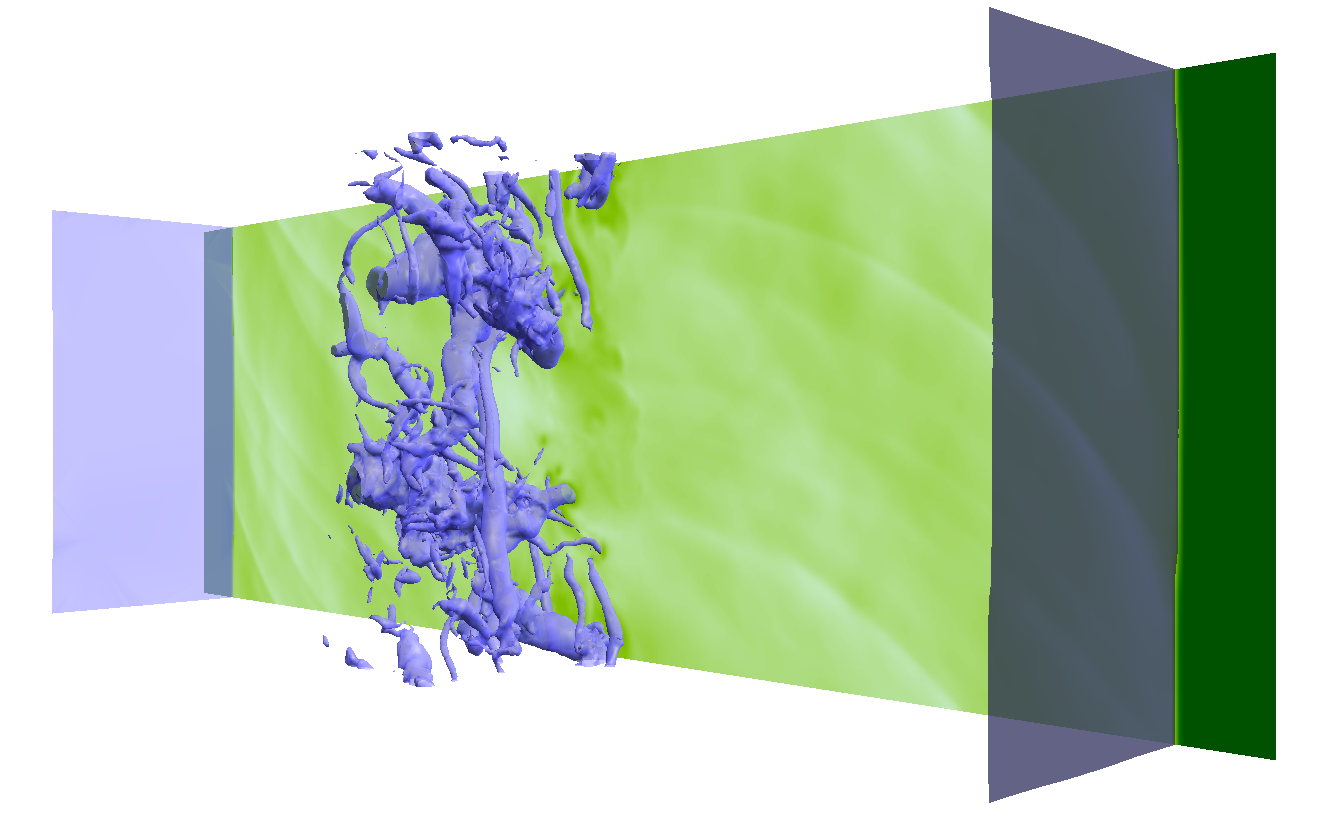}
\caption{Three dimensional distribution of conserved energy density $e$ for problem (d). Subsequent snapshots correspond to evolution times $t = 0.9, 1.8, 2.7, 3.6, 4.5$.}
\label{prob_d_3d}
\end{figure}

\begin{figure}[t!]
\includegraphics[width=6.6cm]{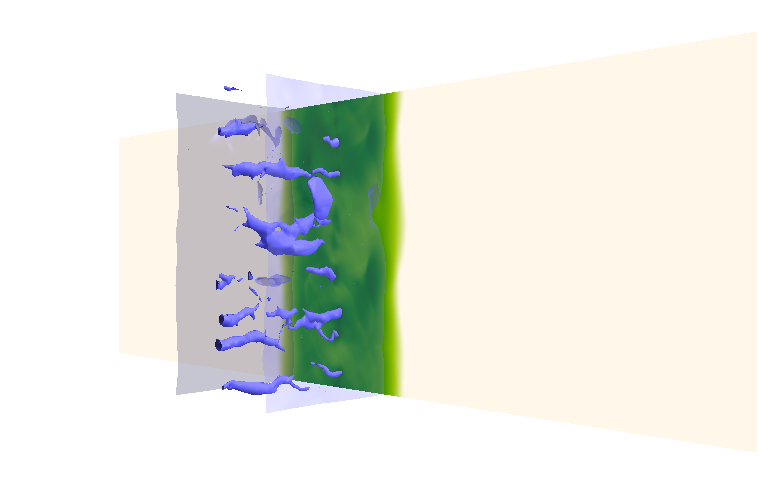}
\includegraphics[width=6.6cm]{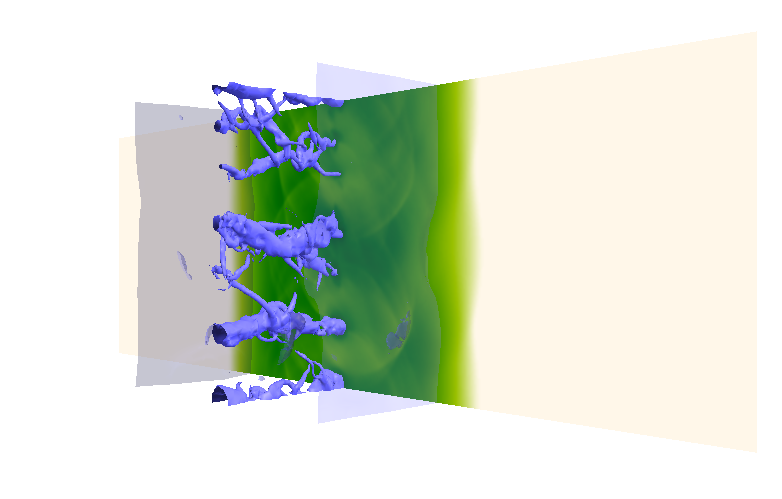}
\includegraphics[width=6.6cm]{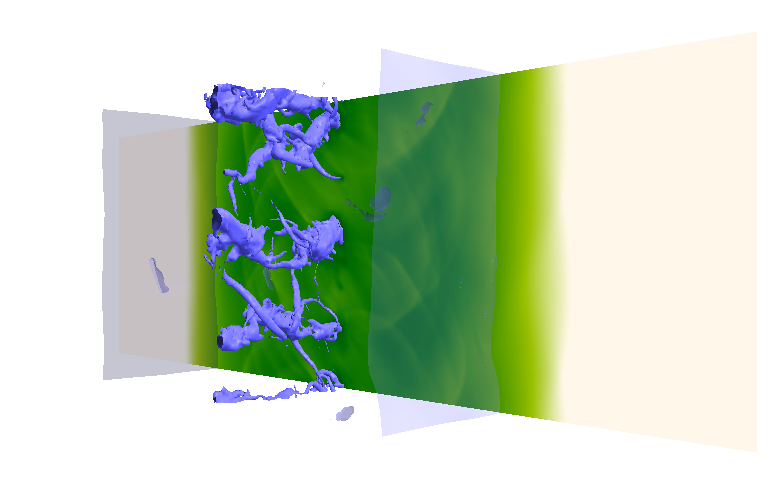}
\includegraphics[width=6.6cm]{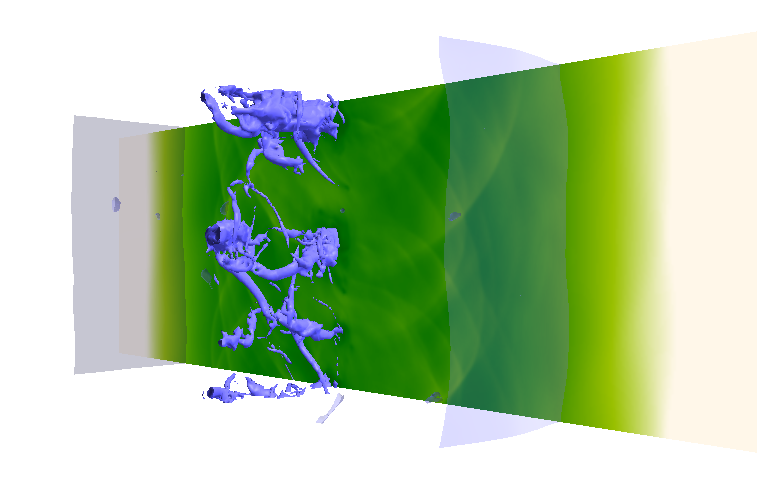}
\includegraphics[width=6.6cm]{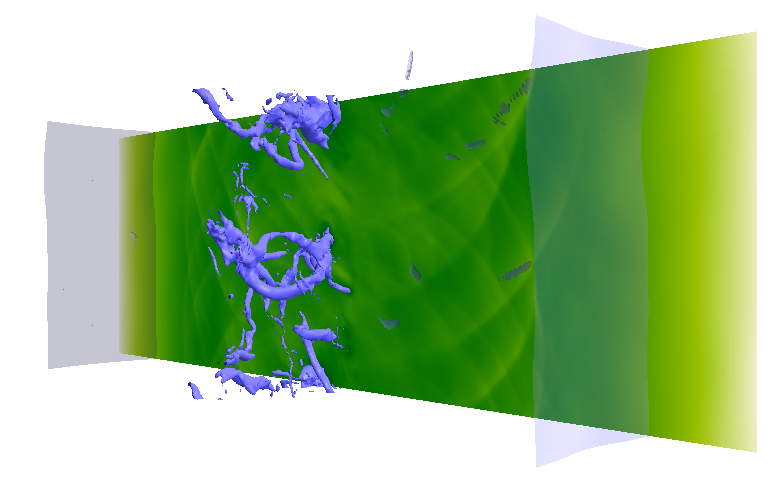}
\caption{Three dimensional distribution of conserved energy density $e$ for problem (e). Subsequent snapshots correspond to evolution times $t = 0.6, 1.2, 1.8, 2.4, 3.0$.}
\label{prob_e_3d}
\end{figure}

\begin{figure}[t!]
\includegraphics[width=6.6cm]{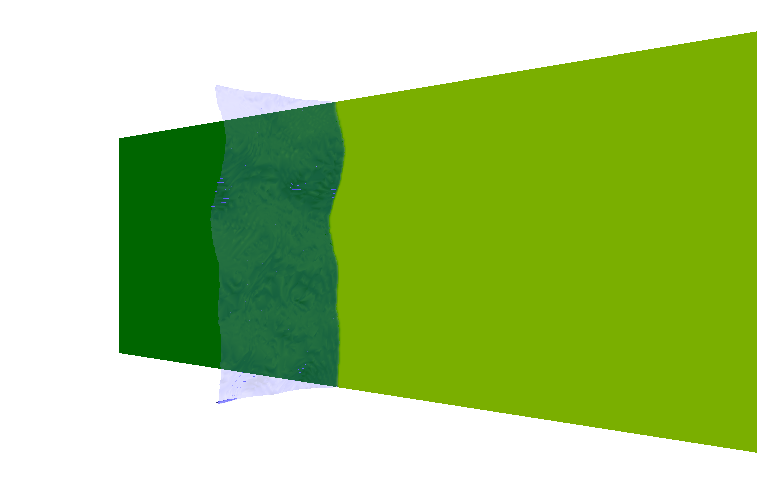}
\includegraphics[width=6.6cm]{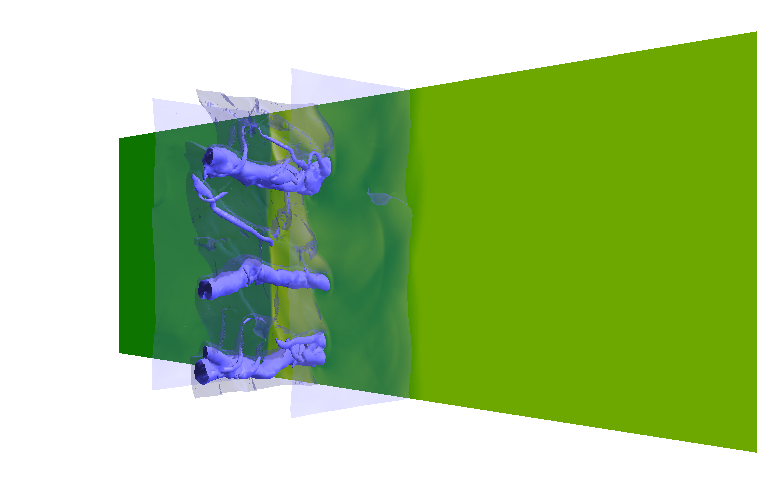}
\includegraphics[width=6.6cm]{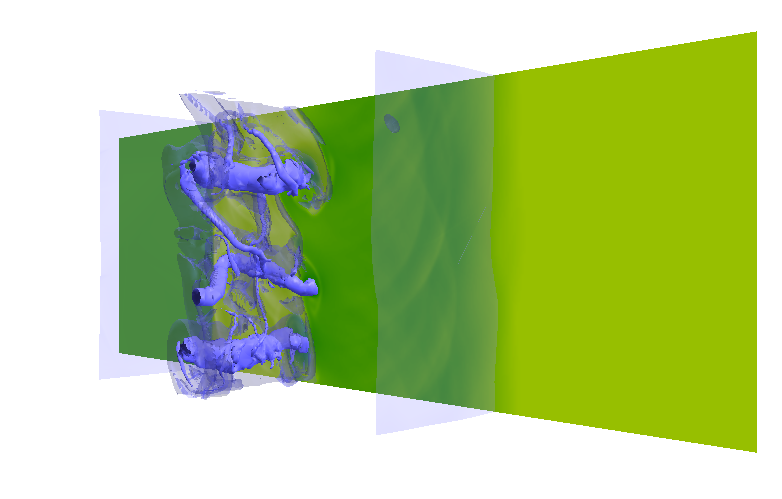}
\includegraphics[width=6.6cm]{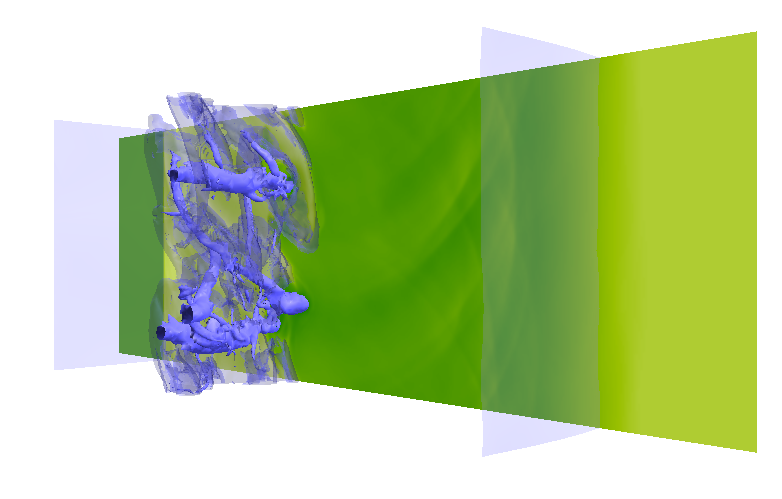}
\includegraphics[width=6.6cm]{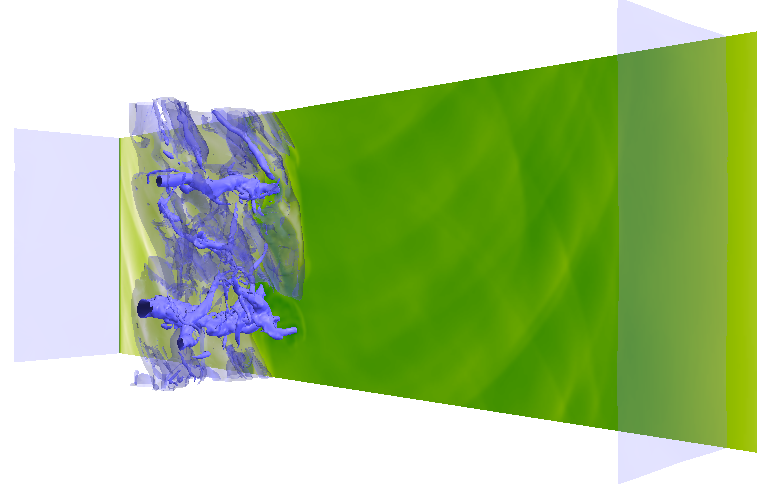}
\caption{Three dimensional distribution of conserved energy density $e$ for problem (f). Subsequent snapshots correspond to evolution times $t = 0.0, 0.9, 1.8, 2.7, 3.6$.}
\label{prob_f_3d}
\end{figure}

\begin{figure}[t!]
\includegraphics[width=8.5cm]{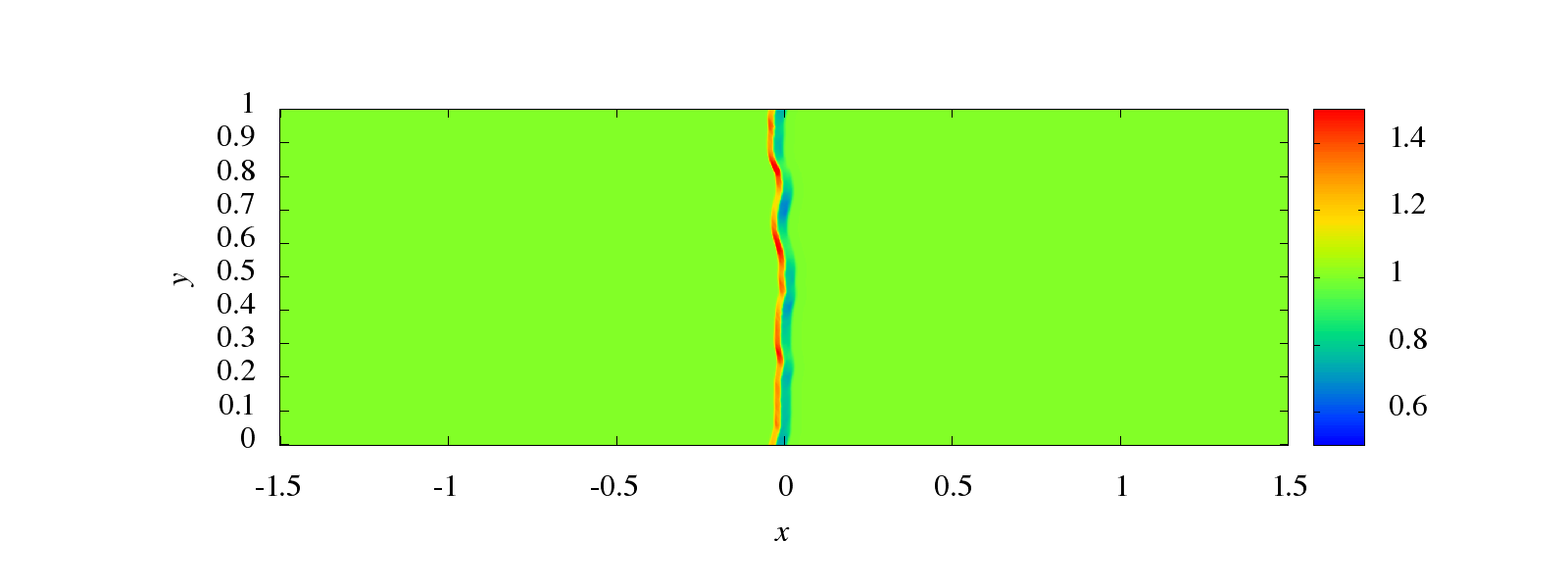}
\includegraphics[width=8.5cm]{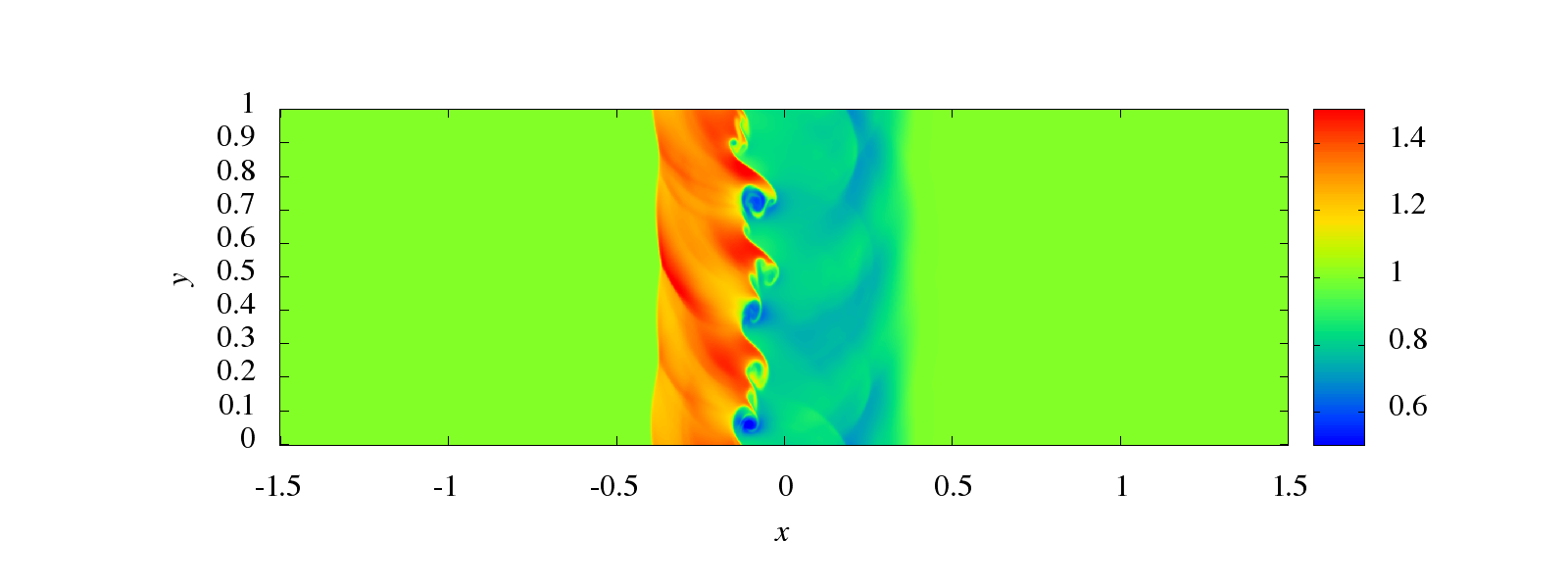}
\includegraphics[width=8.5cm]{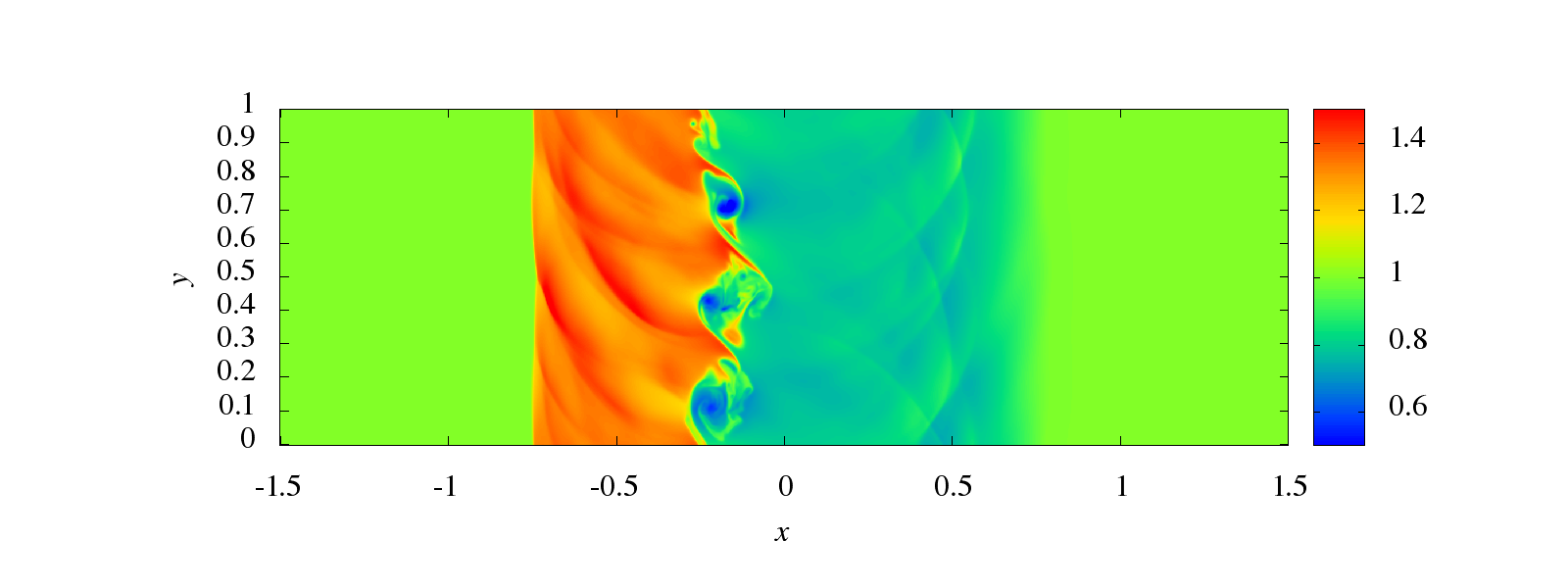}
\includegraphics[width=8.5cm]{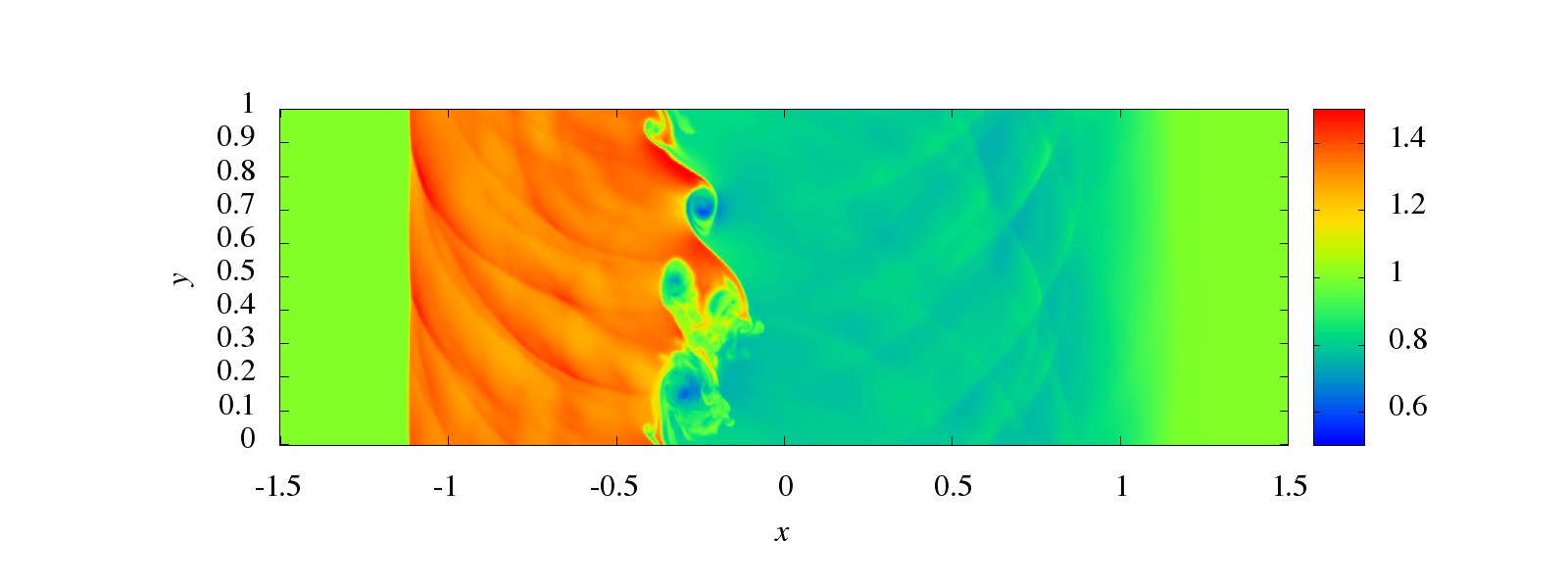}
\includegraphics[width=8.5cm]{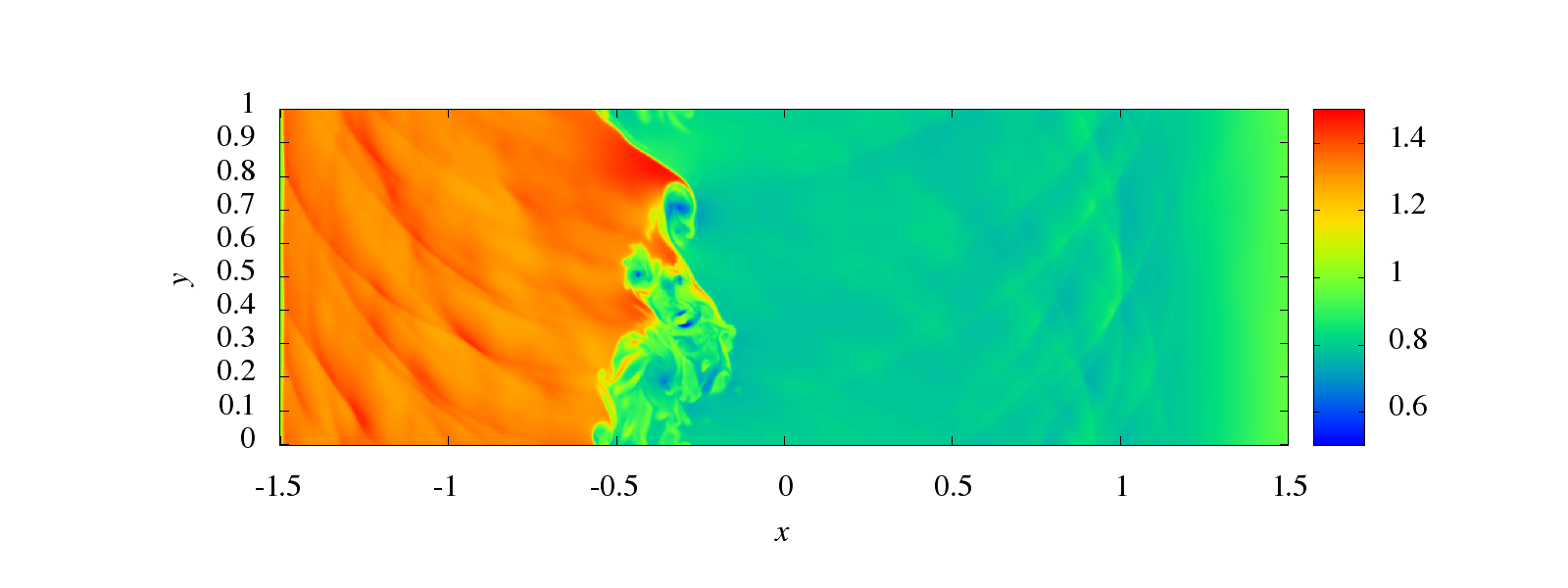}
\caption{Evolution of the rest-mass density $n$ for problem (f). The plots show two dimensional $z = \mathrm{const}$ cross sections through the grid. Subsequent snapshots correspond to evolution times $t = 0.06, 0.9, 1.8, 2.7, 3.6$.}
\label{prob_f_2d}
\end{figure}

\begin{figure}[t!]
\includegraphics[width=8.5cm]{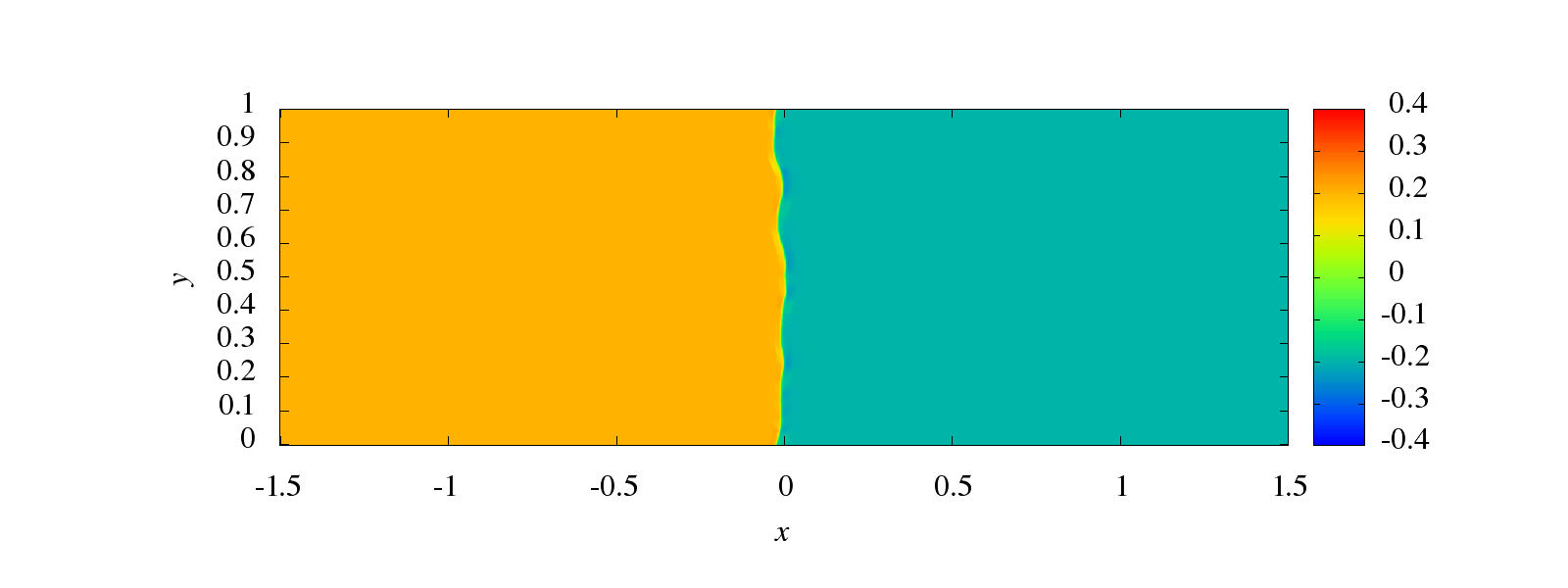}
\includegraphics[width=8.5cm]{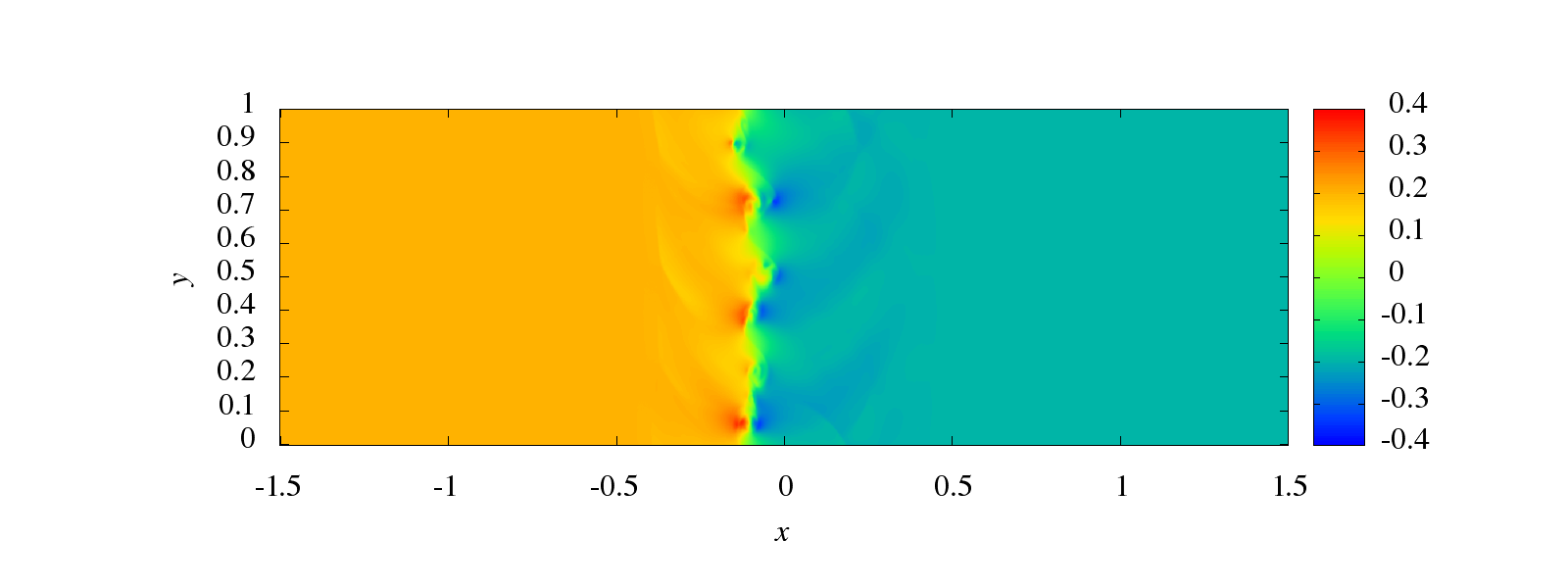}
\includegraphics[width=8.5cm]{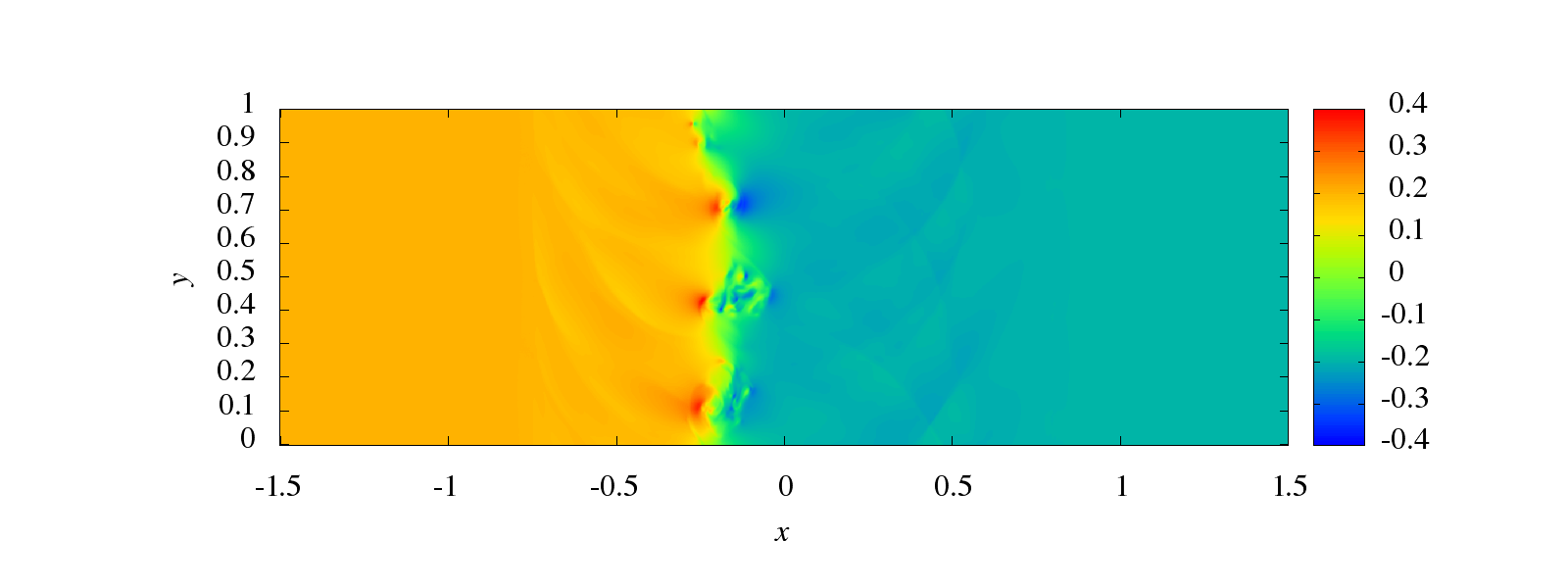}
\includegraphics[width=8.5cm]{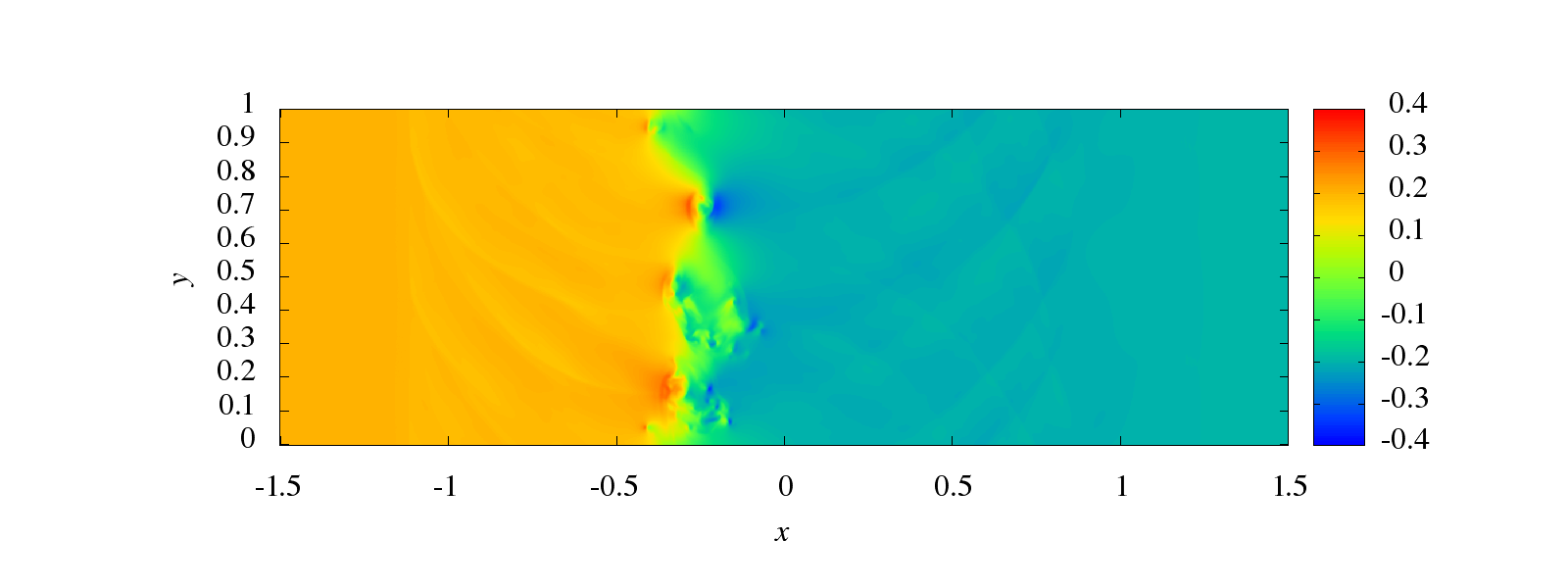}
\includegraphics[width=8.5cm]{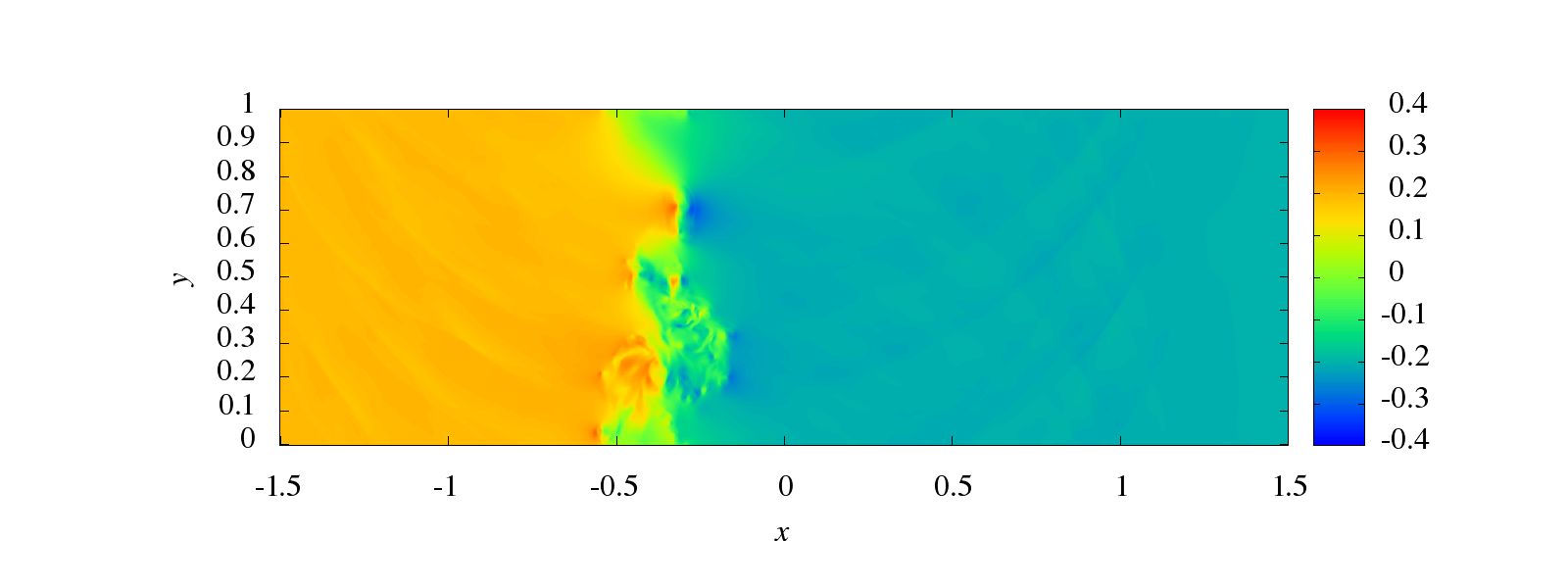}
\caption{Evolution of the $v^y$ component of the velocity for problem (f). The plots show two dimensional $z = \mathrm{const}$ cross sections through the grid. Subsequent snapshots correspond to evolution times $t = 0.06, 0.9, 1.8, 2.7, 3.6$.}
\label{vely_f_2d}
\end{figure}

In this paper we investigate the stability of the relativistic Riemann problem with non-zero tangent velocities. We will, however, restrict ourselves to corrugation instabilities only---perturbations of the initial data will be applied to the shape of the surface dividing two initial states.

Stability of simple waves in the solutions will be first understood according to a very simple notion proposed by Anile and Russo in \cite{anile_russo}. Their idea can be summarized as follows: consider a corrugated shock wave propagating into some medium. If a convex part of the perturbed shock wave moves with a speed that is larger as compared to the speed of an unperturbed shock wave, and reciprocally, if a concave part propagates slower than an unperturbed shock wave, then the perturbations would tend to amplify, and we would say that the wave is unstable. A converse situation would lead to a smoothing of the shock wave and a decrease of the size of corrugations. Such behavior is called stable.

This clearly geometrical notion of stability can be accompanied with some global quantitative data. In order to measure the size of evolving perturbations we have compared each perturbed solution with a corresponding unperturbed one (i.e., a solution for which the initial discontinuity is a plane surface and both initial states are the same as in the original solution). We have decided to look at the perturbation in the conserved energy $e = (\rho + p) W^2 - p$. Technically, we introduce the following three $L$ norms:
\begin{eqnarray}
\label{norms}
\left\Vert e - e_\mathrm{unperturbed}  \right\Vert_{L^1} & = & \int d^3x | e - e_\mathrm{unperturbed} |, \nonumber \\
\left\Vert e - e_\mathrm{unperturbed}  \right\Vert_{L^2} & = & \sqrt{\int d^3x | e - e_\mathrm{unperturbed} |^2}, \nonumber \\
\left\Vert e - e_\mathrm{unperturbed}  \right\Vert_{L^\infty} & = & \mathrm{sup}| e - e_\mathrm{unperturbed} |,
\end{eqnarray}
and we will be interested in their evolution in time.

The Riemann problem is evolved numerically on a 3-dimensional Cartesian grid of $1800 \times 600 \times 300$ zones spanning a cuboid region $x \in [-1.5, 1.5]$, $y \in [0,1]$, $z \in [0,0.5]$. The initial discontinuity is located at the center of the grid, and, modulo perturbations, it is directed perpendicular to the $x$ axis.

On the boundaries in directions $y$ and $z$, that is on the faces perpendicular to the initial discontinuity, we assume periodic boundary conditions. On the remaining two faces, that is in $x$ direction, outflow boundaries are implemented.

Perturbations of the surface of initial discontinuity are applied according to the following formula
\[ f_R(r) = \left\{ \begin{array}{ll} \cos(\frac{\pi r}{2R}), & r \leq R, \\ 0, & r > R, \end{array}  \right.\]
\[ x(y,z) = \sum_i A_i f_{R_i} \left( \sqrt{(x - \bar x_i)^2 + (y - \bar y_i)^2} \right). \]
Here $A_i$, $R_i$, $\bar x_i$, $\bar y_i$ are random amplitudes, radii, and coordinates of the center for each particular perturbation. A little care is required in order to make those perturbation compatible with the assumption of periodicity at the boundaries.

We decided to focus on a set of six solutions (a)--(f) with moderately relativistic fluid velocities ($v \sim 1/2$), but exhausting all possible different wave patterns for systems I and II. Values characterizing initial states for these problems are collected in Table 1. System I was evolved with $c_s^2 = 1/3$. For system II we have chosen $\gamma = 4/3$. This corresponds to equations of state $p = \rho/3$ and $p = n \epsilon /3$ respectively.

\begin{figure}[t!]
\includegraphics[width=7.5cm]{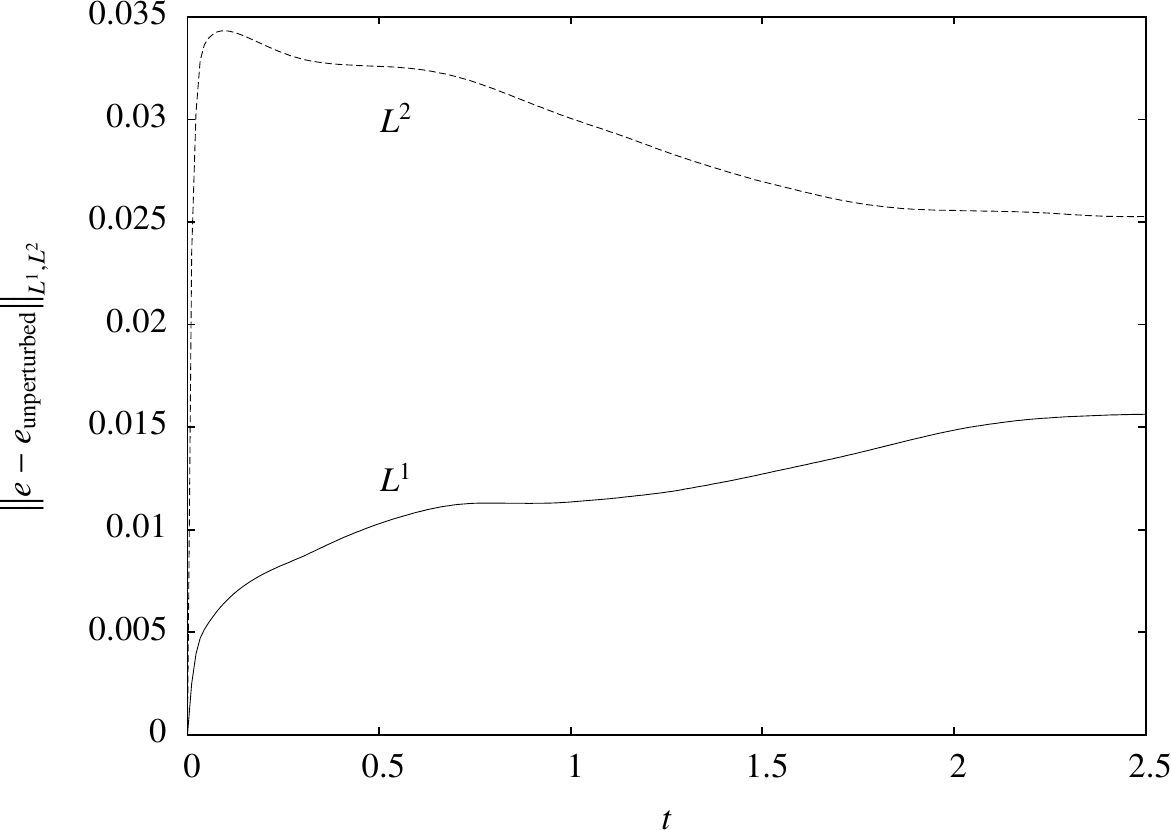}
\caption{The growth of the perturbation in the energy $e$ computed with respect to $L^1$ and $L^2$ norms (Eqs.~(\ref{norms})) for problem (a).}
\label{normsa}
\end{figure}

\begin{figure}[t!]
\includegraphics[width=7.5cm]{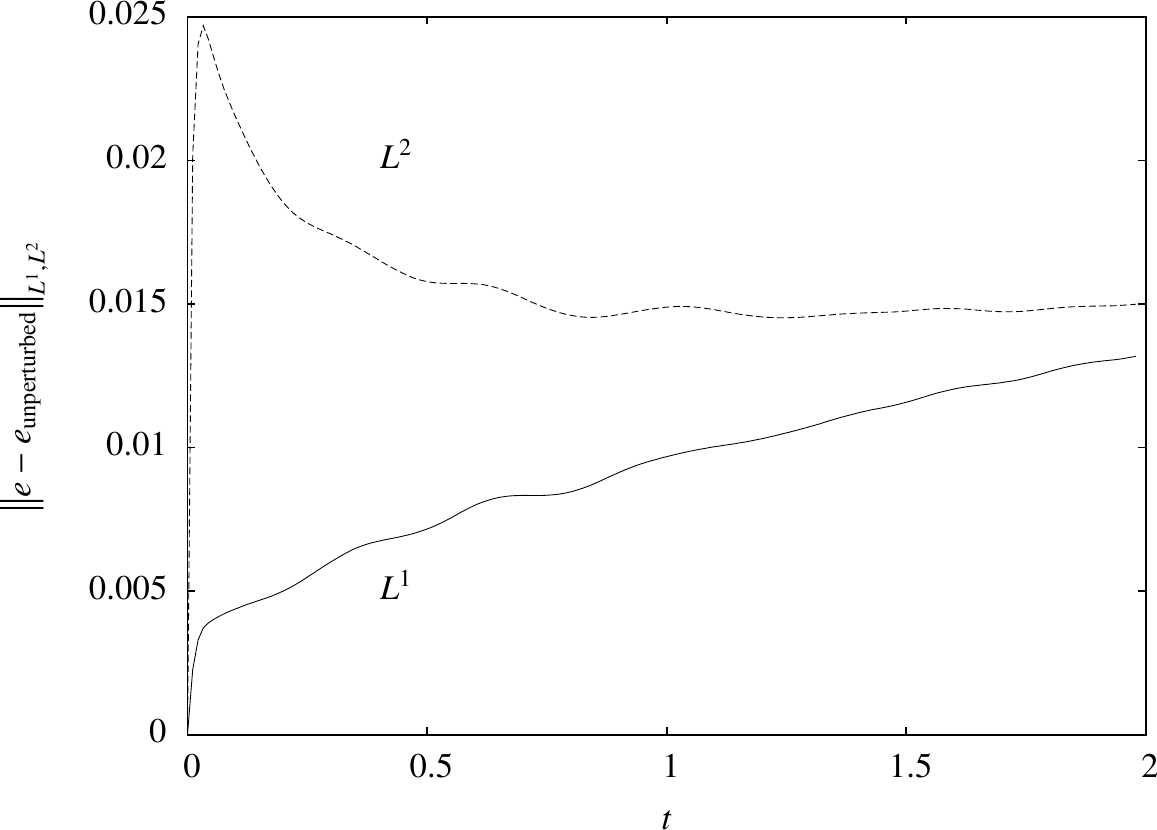}
\caption{The growth of the perturbation in the energy $e$ computed with respect to $L^1$ and $L^2$ norms (Eqs.~(\ref{norms})) for problem (b).}
\label{normsb}
\end{figure}

\begin{figure}[t!]
\includegraphics[width=7.5cm]{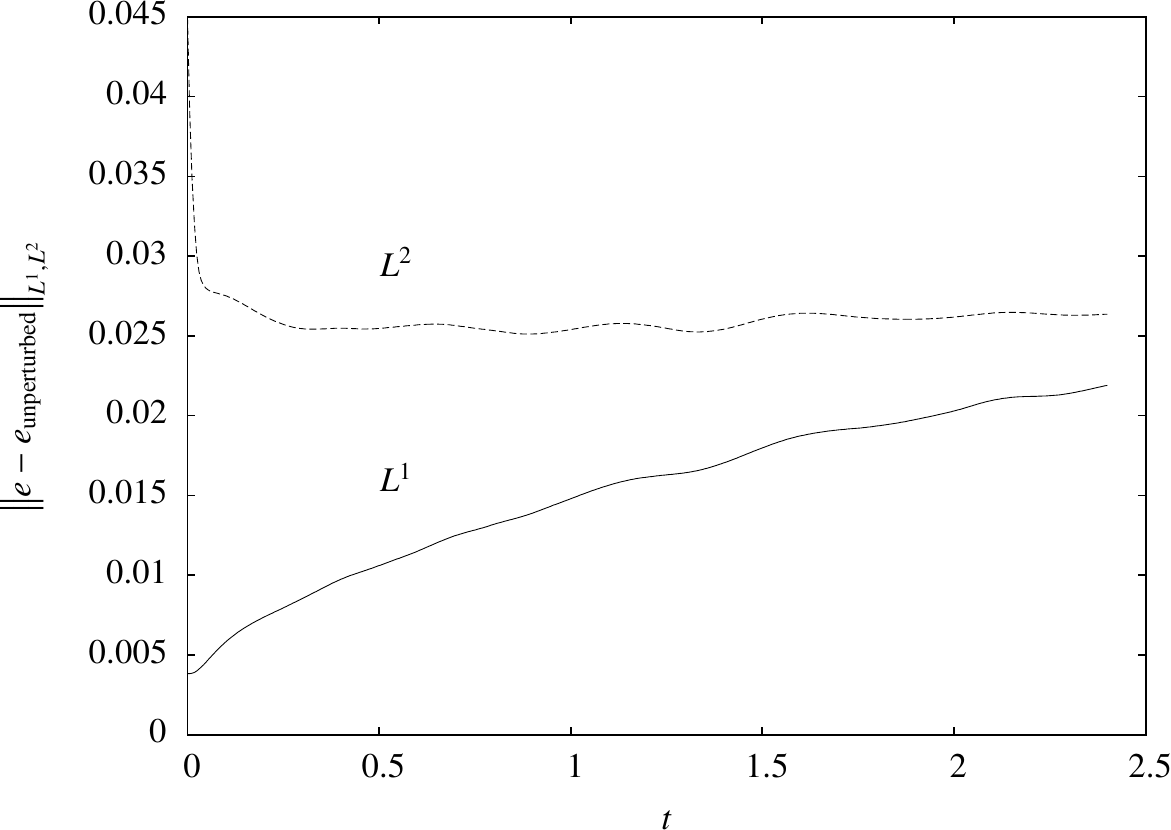}
\caption{The growth of the perturbation in the energy $e$ computed with respect to $L^1$ and $L^2$ norms (Eqs.~(\ref{norms})) for problem (c).}
\label{normsc}
\end{figure}

\begin{figure}[t!]
\includegraphics[width=7.5cm]{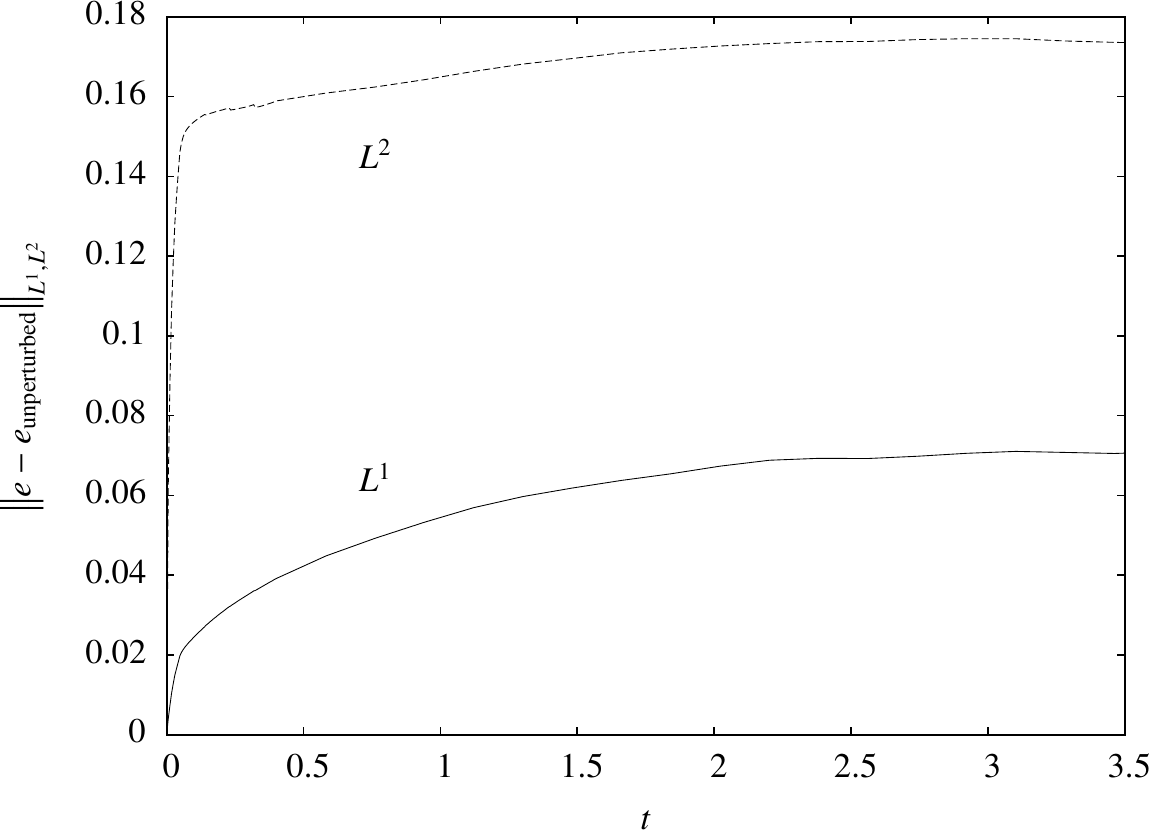}
\caption{The growth of the perturbation in the energy $e$ computed with respect to $L^1$ and $L^2$ norms (Eqs.~(\ref{norms})) for problem (d).}
\label{normsd}
\end{figure}

\begin{figure}[t!]
\includegraphics[width=7.5cm]{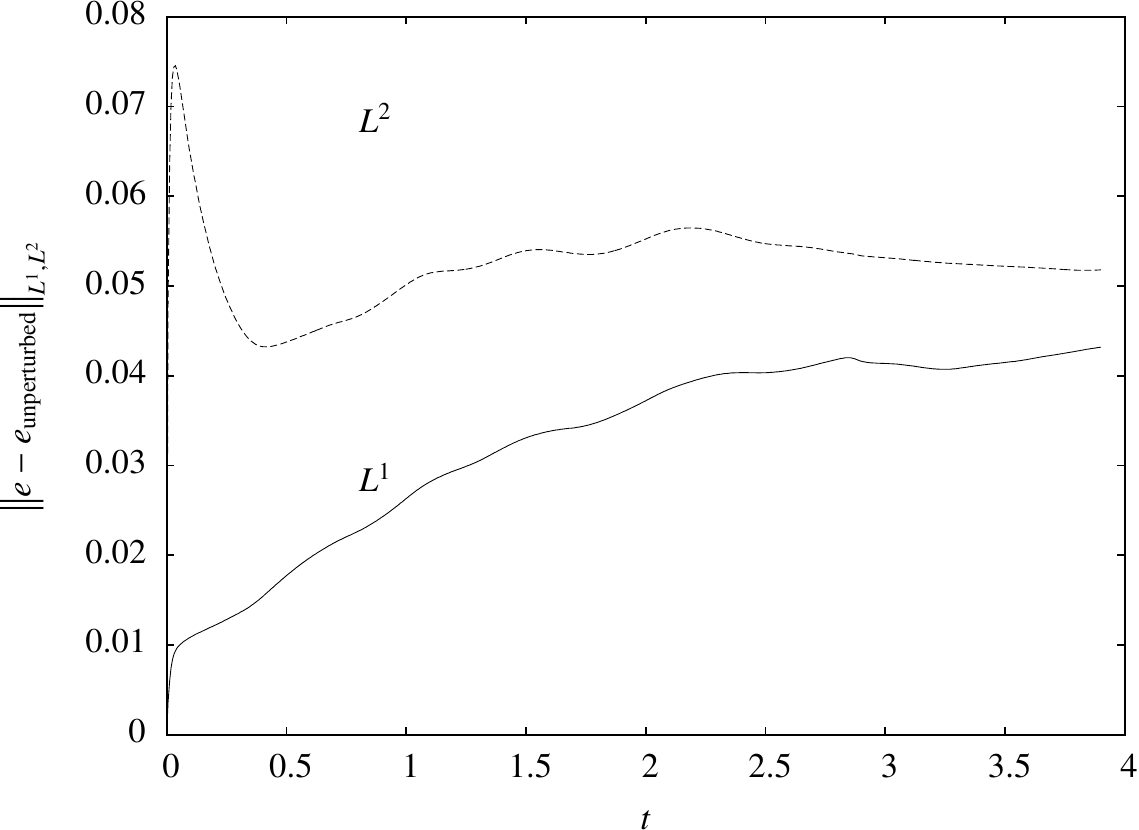}
\caption{The growth of the perturbation in the energy $e$ computed with respect to $L^1$ and $L^2$ norms (Eqs.~(\ref{norms})) for problem (e).}
\label{normse}
\end{figure}

\begin{figure}[t!]
\includegraphics[width=7.5cm]{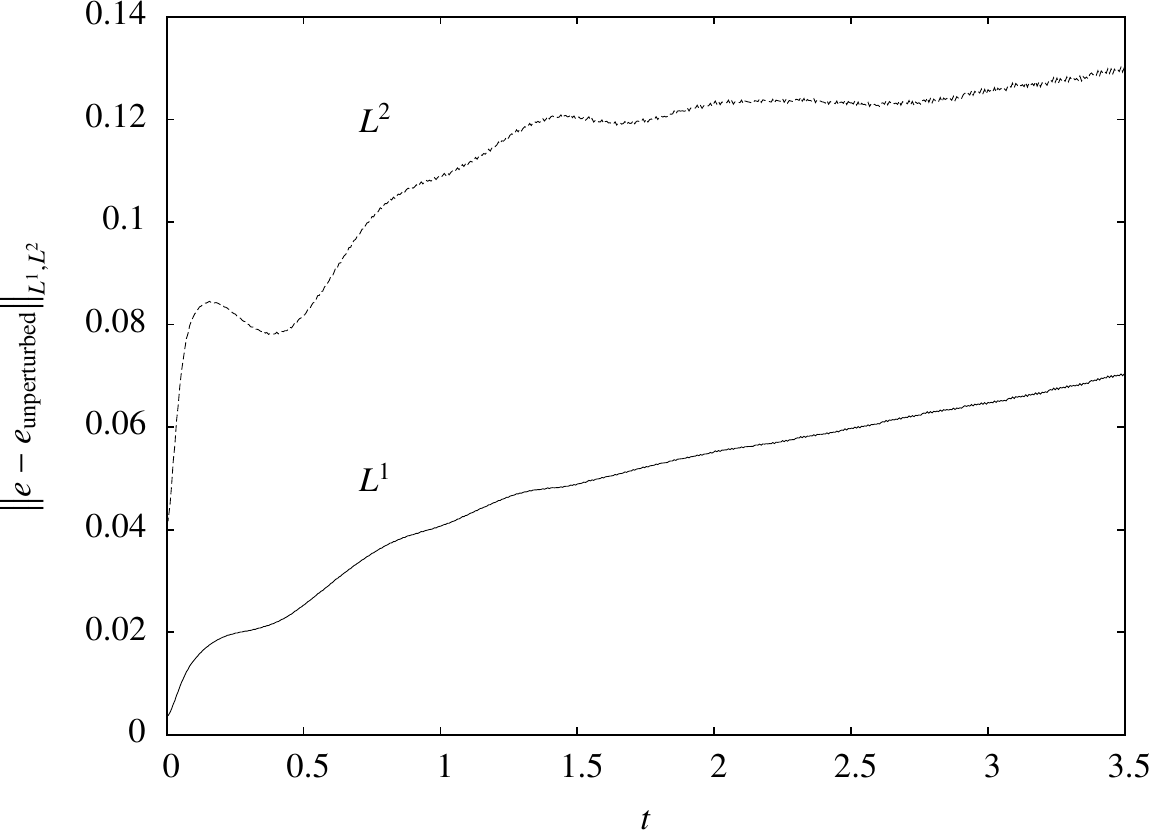}
\caption{The growth of the perturbation in the energy $e$ computed with respect to $L^1$ and $L^2$ norms (Eqs.~(\ref{norms})) for problem (f).}
\label{normsf}
\end{figure}

\begin{figure}[t!]
\includegraphics[width=7.5cm]{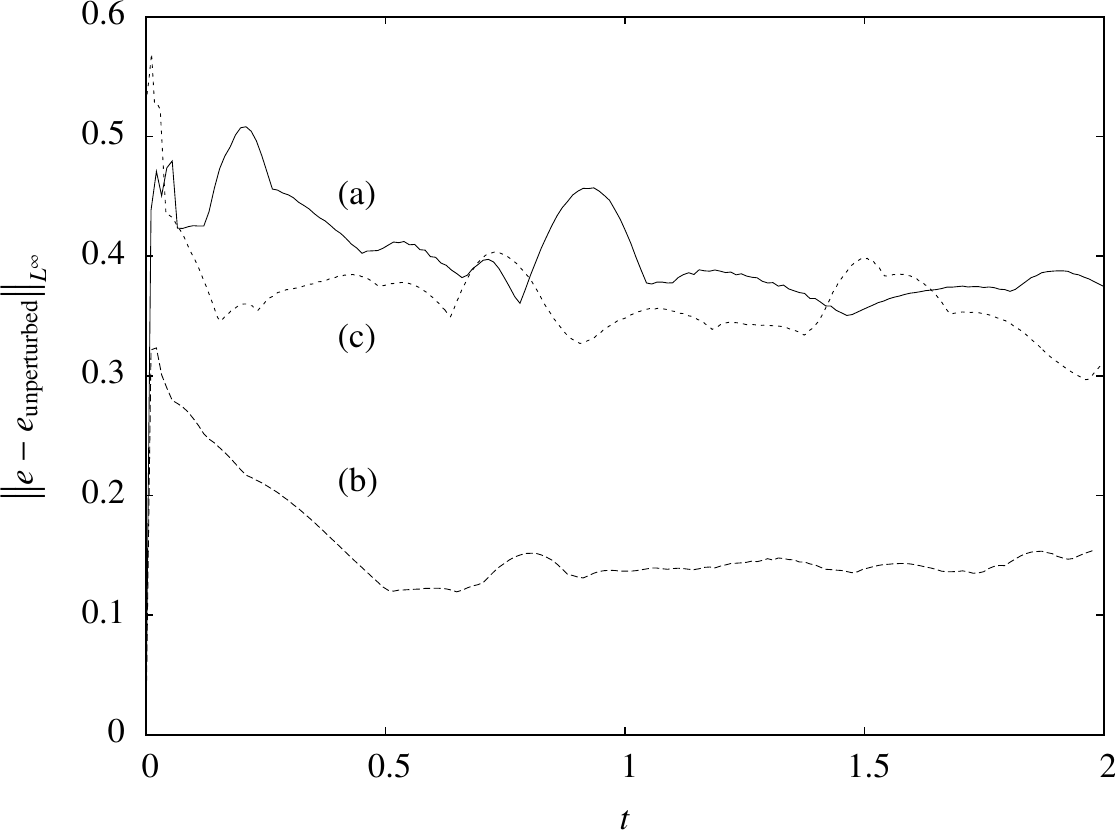}
\caption{The growth of the perturbation in the energy $e$ computed with respect to $L^\infty$ norm (Eqs.~(\ref{norms})) for problems (a), (b) and (c).}
\label{norms_inf}
\end{figure}

\begin{figure}[t!]
\includegraphics[width=7.5cm]{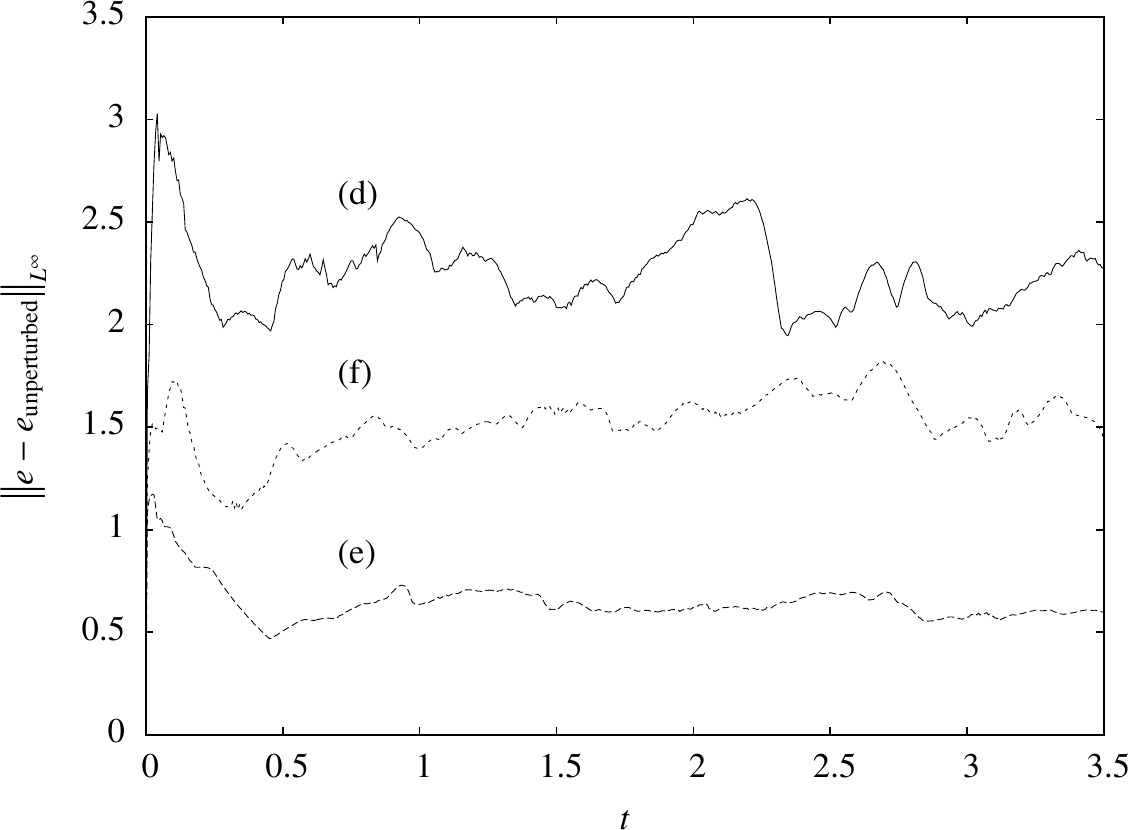}
\caption{The growth of the perturbation in the energy $e$ computed with respect to $L^\infty$ norm (Eqs.~(\ref{norms})) for problems (d), (e) and (f).}
\label{norms_inf2}
\end{figure}

Graphs showing the evolution of the initial data for all problems (a)--(f) are presented on Figs.~\ref{prob_a_3d}--\ref{prob_f_3d}. In each case a series of three dimensional perspective plots corresponding to subsequent moments in time is shown. The surfaces on the plots are the surfaces of constant energy density $e$ (isopycnic surfaces). Also, on the back face of the grid the boundary values of the energy density $e$ are color-coded. All pictures are oriented in such a way that the $x$ axis is pointing slightly in the direction of the viewer.

The common feature that can be observed on all plots is the existence of an unstable turbulent region around the contact discontinuity with instabilities in the form of twisted tubes of considerably lower density. Another common feature is that in all cases both shock waves and rarefaction waves behave in a stable way---their shape gets smoother and smoother with time. One can notice, however, that the behavior of problems with the perfect gas equation of state is more turbulent than that of problems with the ultrarelativistic equation of state.

The instability occurring around the contact discontinuity can be identified as a relativistic version of the Kelvin--Helmholtz instability. In all cases (a)--(f) the gas in the intermediate states $L_\ast$ and $R_\ast$ flows with some nonzero tangent velocity $v^y$, different at both sides of the contact discontinuity. We have checked that in the absence of the tangent velocity no instability around the contact discontinuity develops. The Kelvin--Helmholtz character of the instability can be seen on Fig.~\ref{prob_f_2d}, presenting once again the evolution of problem (f). Here, we have decided to show the distribution of the rest-mass density $n$ on two dimensional $z = \mathrm{const}$ cross sections through the grid. In this case the two intermediate states $L_\ast$ and $R_\ast$ have different densities and we can observe a standard Kelvin--Helmholtz wave picture. The tubes of lower density can be identified with regions trapped under crests of the waves.

In the same fashion Fig.~\ref{vely_f_2d} shows the two dimensional plots of the $v^y$ component of the velocity for problem (f), i.e., for the same data as those presented on Fig.~\ref{prob_f_2d}.

Figs.~\ref{normsa}--\ref{norms_inf2} show the time evolution of norms (\ref{norms}) for solutions (a)--(f). The evolution of $L^1$ and $L^2$ norms for those problems is depicted on Figs.~\ref{normsa}--\ref{normsf}. In both norms there is a short phase of a rapid growth of the perturbation followed by either a decrease of its size ($L^2$ norm) or a very slow growth ($L^1$ norm). For completeness $L^\infty$ norms for problems (a)--(f) are shown on Figs.~\ref{norms_inf} and \ref{norms_inf2}. Also this norm increases at early stages of evolution, and starts to oscillate around some constant value for later times. Judging from those results, one can say that the fast exponential growth of instabilities, usually obtained in a linearized calculation for the non relativistic Kelvin--Helmholtz instability, is only limited to a very short early phase of evolution. Later, nonlinear effects become dominant, and the growth of perturbations is stalled.

\section{Summary}

We have performed a series of three dimensional numerical studies of the corrugation stability of the Riemann problem in relativistic hydrodynamics with non-zero velocities tangent to the surface of the initial discontinuity. We specialized to two equations of state: the ultrarelativistic one and that of perfect gas. In both cases a modern high resolution shock capturing Godunov type numerical scheme was employed. The conserved hydrodynamical quantities were evolved in time within the framework of method of lines, using a standard fourth order Runge--Kutta algorithm. Numerical fluxes were based on the work of Donat and Marquina \cite{marquina}. The required spectral decomposition of the Jacobians appearing in the equations of relativistic hydrodynamics was taken from \cite{banylus_et_al} for the perfect gas equation of state; it was computed separately for the general barotropic case (ultrarelativistic equation of state, in particular), and the results are presented in this paper. The CENO reconstruction procedure was adapted from \cite{liu_osher,londrillo_zanna,zanna_bucciantini}.

We have focused on mildly relativistic solutions resulting with different possible wave patterns, i.e., solutions with two shock waves, two rarefaction waves, or a combination of a shock wave and a rarefaction wave. In all cases both rarefaction and shock waves behaved in a stable way in the sense that their shapes were becoming flattened with time. Kelvin--Helmholtz type instabilities were observed to develop only around the contact discontinuity, forming a turbulent region with characteristic rarefaction ``tubes''. This behavior is essentially the same for the perfect gas equation of state and for the ultrarelativistic one. Slightly more turbulent instabilities develop in the case of perfect gas equation of state, what can be observed both on the three dimensional plots of solutions and in the evolution of norms measuring the size of the perturbation. The behavior of perturbed solutions with ultrarelativistic velocities tangent to the initial discontinuity remains an open question.

We believe that these results can be helpful in understanding more complex phenomena occurring, for instance, during propagation of relativistic astrophysical jets. The evolution of the so-called turbulent cocoon \cite{marti_et_al} could be compared with our results concerning the Riemann problem alone. Since the behavior of turbulent flows depends on dimensionality, and in particular it is different in two and three spatial dimensions, a fair comparison would require three dimensional simulations of jets (see \cite{mignone_et_al} for an example).

Solutions of the relativistic Riemann problems are also applied in the modelling of heavy ion collisions during the so-called hydrodynamical phase (cf.~\cite{bouras_et_al} and references therein). Our results should be of interest here.

\acknowledgments
I wish to thank Dr.~Andrzej Odrzywo\l{}ek for many fruitful discussions. He was also a careful reader of the manuscript of this paper.
 
The research was carried out with the supercomputer ``Deszno'' purchased thanks to the financial support of the European Regional Development Fund in the framework of the Polish Innovation Economy Operational Program (contract no. POIG.02.01.00-12-023/08).

%%%%%%%%%%%%%%%%%%%%%%%%%%%%%%%%%%%%%%%%%%%%%%%%%%%%%%%%%%%%%%%%%%%%%%%%%%%%%%%%
%%%%%%%%%%%%%%%%%%%%%%%%%%%%%%%%%%%%%%%%%%%%%%%%%%%%%%%%%%%%%%%%%%%%%%%%%%%%%%%%
%%%%%%%%%%%%%%%%%%%%%%%%%%%%%%%%%%%%%%%%%%%%%%%%%%%%%%%%%%%%%%%%%%%%%%%%%%%%%%%%

\end{document}